\documentclass{article}

\usepackage{color,graphics,epsfig,rotating}

\def\greaterthansquiggle{\raise.3ex\hbox{$>$\kern-.75em\lower1ex\hbox{$\sim$}}}
\def\lessthansquiggle{\raise.3ex\hbox{$<$\kern-.75em\lower1ex\hbox{$\sim$}}}
\newcommand{\beq}{\begin{equation}}
\newcommand{\eeq}{\end{equation}}
\newcommand{\beqa}{\begin{eqnarray}}
\newcommand{\eeqa}{\end{eqnarray}}
\newcommand{\ba}{\begin{array}}
\newcommand{\ea}{\end{array}}

\newcommand{\grts}{\greaterthansquiggle}
\newcommand{\lets}{\lessthansquiggle}

\newcommand{\ra}{\rightarrow}

\def\a               {\alpha}

\def\ti    {\tilde}

\def\snu   {{\ti\nu}}

\newcommand{\T}{\tilde\tau}
\newcommand{\SN}{\tilde\nu_{\tau}}
\newcommand{\CH}{\tilde\chi}
\newcommand{\CPM}{\cos\varphi_{\mu}}
\newcommand{\h}{h^{\tau}}
\newcommand{\f}{f^{\tau}}

\newcommand{\gsim}{\;\raisebox{-0.9ex}
           {$\textstyle\stackrel{\textstyle >}{\sim}$}\;}
\newcommand{\lsim}{\;\raisebox{-0.9ex}{$\textstyle\stackrel{\textstyle <}
           {\sim}$}\;}

\begin{document}  
\setlength{\unitlength}{1mm}

\begin{flushright}
  UWThPh-2001-36\\
  HEPHY-PUB 742 \\
  TGU-28\\
  ZU-TH 39/01\\
  IFIC/02-30\\
  hep-ph/0207186\\[3mm]
\end{flushright}

\begin{center}

{\Large \bf\boldmath $\tau$-Sleptons and
$\tau$-Sneutrino in the MSSM with Complex Parameters}

\vspace{5mm}

{\large A.~Bartl,$^1$~ 
K.~Hidaka,$^2$~
T.~Kernreiter,$^{1,3}$~
W.~Porod\,$^{4,5}$} \\

\vspace{4mm}

{\normalsize \it
$^1$~Institut f\"ur Theoretische Physik, Universit\"at Wien, 
     A--1090 Vienna, Austria \\
$^2$~Dept. of Physics, Tokyo Gakugei University, Koganei, 
     Tokyo 184--8501, Japan \\
$^3$~Instituto de Fisica Corpuscular-C.S.I.C./Universitat de
Valencia, E-46071 Valencia, Spain \\
$^4$~Inst.~f.~Hochenergiephysik, \"Oster.~Akademie d.~Wissenschaften,
      A-1050 Vienna, Austria \\
$^5$~Inst.~f\"ur Theor. Physik, Universit\"at Z\"urich, CH-8057 Z\"urich,
      Switzerland}

\end{center}

\begin{abstract} 
We present a phenomenological study of $\tau$-sleptons $\T_{1,2}$ and 
$\tau$-sneutrinos $\SN$ in the Minimal Supersymmetric
Standard Model with complex parameters $A_{\tau}$,
$\mu$ and $M_1$. We analyse production and decays
of the $\T_{1,2}$ and $\SN$ at a future $e^+e^-$ collider.
We present
numerical predictions for the important decay rates, paying
particular attention to their dependence on the complex
parameters. The branching ratios of the fermionic decays of
$\ti\tau_1$ and $\ti \nu_{\tau}$ 
show a significant phase dependence for 
$\tan\beta \lets 10$. For $\tan\beta \grts 10$ the branching
ratios for the $\T_2$ decays into Higgs bosons depend very 
sensitively on the phases. We show how information on
the phase $\varphi_{A_{\tau}}$ and the other fundamental $\T_i$ 
parameters can be obtained from measurements of 
the $\T_i$ masses, polarized cross sections 
and bosonic and fermionic decay branching ratios,
for small and large $\tan\beta$ values. We estimate  the expected errors 
of these parameters. 
Given favorable conditions, the error of
$A_\tau$ is about 10\% to 20\%, while the errors of the remaining stau
parameters are in the range of approximately 1\% to 3\%.
We also show
that the induced electric dipole moment of the  $\tau$--lepton 
 is well below the current experimental limit.
\end{abstract}

\section{Introduction}
So far most phenomenological studies on supersymmetric (SUSY)
particle searches 
have been performed within the Minimal Supersymmetric Standard
Model (MSSM) with real SUSY parameters. In this paper we study
the production and decays of $\tau$-sleptons and
$\tau$-sneutrinos at an $e^+e^-$ linear collider in the MSSM
with complex SUSY parameters.

In the SUSY extension of the Standard Model
(SM) one introduces scalar leptons $\ti \ell_L$, 
$\ti \ell_R$, scalar neutrinos $\ti \nu_{\ell}$ and scalar
quarks $\ti q_L$, $\ti q_R$ as the
SUSY partners of the leptons $\ell_{L,R}$, 
neutrinos $\nu_{\ell}$ and quarks $q_{L,R}$, respectively
\cite{nilles}. For each definite fermion flavor 
the states $\ti f_L$ and $\ti f_R$ are mixed by 
Yukawa terms. The mass eigenstates are $\ti f_1$ and
$\ti f_2$, with $m_{\ti f_1} < m_{\ti f_2}$ \cite{elru}. 
For the sfermions of the first and second generation 
$\ti f_L - \ti f_R$ mixing can be neglected. For the third
generation sfermions, however, $\ti f_L - \ti f_R$ mixing has 
to be taken into account due to the larger Yukawa
coupling \cite{dreesnojiri,bamapo}. 

In the case of the $\tau$-sleptons $\ti \tau_L - \ti \tau_R$
mixing is important if the SUSY parameter $\tan \beta$ is large,
$\tan \beta \gsim 20$. The lower mass eigenvalue
$m_{\ti \tau_1}$ can be rather small and the $\ti \tau_1$ could
be the lightest charged SUSY particle. 
The experimental search for the $\tau$-sleptons and the
$\tau$-sneutrino and the determination of their parameters is,
therefore, 
an important issue at all present and future colliders. Pair
production of $\tau$-sleptons and $\tau$-sneutrinos will 
be particularly interesting at an $e^+e^-$ linear collider with
centre of mass energy $\sqrt{s} = 0.5-1.2$~TeV. At such a
collider and with an integrated
luminosity of about $500$~fb$^{-1}$ it will be possible
to measure masses, cross sections and decay branching
ratios with high precision \cite{acco,tdr}. 
This will allow us to obtain information on 
the fundamental soft SUSY breaking parameters of the 
third generation slepton system.

In the recent phenomenological study of 3rd generation sfermions
in the real MSSM it has been shown how the masses and the mixing
angle of the stop system can be determined by measurements of
the production cross sections with polarized beams \cite{sferm}.
The results of a simulation of 
$e^+e^- \to \ti t_1 \bar{\ti t_1}$ with the decay modes 
$\ti t_1 \to \CH^0_1 c$ and $\ti t_1 \to \CH^+_1 b$ and
including full SM background in \cite{sop} imply that 
with an integrated luminosity of $500$~fb$^{-1}$ an accuracy of
the order of $1\%$ or better may be obtained. The numerical 
precision to be expected for the determination of the underlying
SUSY parameters $M_{\ti Q}$, $M_{\ti U}$ and (real) $A_t$ has
also been given. For low $\tan\beta$ one can expect similar
results for the sbottom and stau systems \cite{acco,tdr,sferm}. 

The assumption of real SUSY parameters has partly been justified
by the very small experimental upper limits on the electric
dipole moments (EDM) of electron and neutron. A
possibility to avoid the EDM constraints is to assume that the
masses of the first and second generation sfermions are large
(above the TeV scale), while the masses of the third
generation sfermions are small (below $1$~TeV) \cite{cohen}.
Another possibility is suggested by recent analyses of the
 EDMs, which have shown that strong
cancellations between the 
different SUSY contributions to the EDMs can occur \cite{nath}.
As a consequence of these cancellations it has turned out that
the complex phase of the Higgs--higgsino mass parameter $\mu$ 
is much less restricted than previously assumed, whereas the
complex phases of the soft--breaking trilinear scalar 
coupling parameters $A_f$ are practically
unconstrained \cite{EDM1,EDM2}. For example, in a mSUGRA--type
model with universal parameters $M_{1/2}$, $M_0$, $\tan\beta$
and complex $A_0$, with $|\mu|^2$ being determined by radiative 
electroweak symmetry breaking, the phase of $\mu$ is constrained
to $|\varphi_{\mu}| \lets 0.1-0.2$ for low 
values of the scalar mass parameter, $M_0 \lets 400$~GeV, and
becomes less constrained for higher values of $M_0$. The phase
of $A_0$, $\varphi_{A_0}$, turns out to be correlated with
$\varphi_{\mu}$, but otherwise not restricted
\cite{EDM2,bagama}. In models 
with more general parameter specifications also $\varphi_{\mu}$
turns out to be less constrained \cite{kane}.
In any case, this means that in a complete
phenomenological analysis of production and decays of third
generation sfermions one has to take into account that the SUSY
parameters $\mu$ and $A_f$ may be complex and one has to study
the implications that follow for the important observables. 

In our present phenomenological study of 3rd generation sleptons
we use the MSSM as general framework and we assume that the
parameters $\mu$, $A_{\tau}$ and $M_1$ are complex ($A_{\tau}$
is the trilinear scalar coupling parameter of the 
$\ti \tau_i$-system and $M_1$ is the $U(1)$ gaugino mass
parameter). We neglect flavor changing $CP$ violating 
phases and assume that the scalar mass matrices and trilinear
scalar coupling parameters are flavor diagonal.
We perform an analysis of production and decay rates of 
$\ti \tau_1$, $\ti \tau_2$ and $\ti \nu_{\tau}$
at an $e^+e^-$ 
linear collider with a CMS energy $\sqrt{s}=0.5-1.2$~TeV.
We include also explicit $CP$ violation in the Higgs sector
induced by stop and sbottom loops with complex parameters
as in \cite{carenaetal,pilaftsi1} and \cite{dreescp}, 
using the loop--corrected formulae of \cite{carenaetal}.
Our present study is an extension of the corresponding one in
the MSSM with real parameters in \cite{sferm}. 
Compared to the real MSSM, the inclusion of the complex 
phases $\varphi_{A_{\tau}}$, $\varphi_{\mu}$ and
$\varphi_{U(1)}$ 
of $A_{\tau}$, $\mu$ and $M_1$ means that the number of 
independent fundamental SUSY parameters is increased. In order
to determine all these parameters one has to measure more
independent observables than in the real case. 

In principle, the imaginary parts of the complex parameters
involved could most directly and unambiguously be determined by 
measuring suitable $CP$ violating observables. However, in the 
$\ti \tau_i$-system this is not straightforward, because the 
$\ti \tau_i$ are spinless and their main decay modes are
two--body decays. A possible method has been proposed in
\cite{drees1}, which is applicable if the mass splitting between
the mass eigenstates $\ti\tau_1$ and $\ti\tau_2$ is very small.
If $m_{\ti\tau_1} - m_{\ti\tau_2}$ is of the order of the
decay widths, $\T_1 - \T_2$ oscillations will occur which can
lead to large $CP$ violating asymmetries in $e^+e^-$
annihilation. In Ref. \cite{drees2} an analysis of 
$\mu^+ \mu^- \rightarrow \ti\tau_i \bar{\ti\tau_j}$ with
longitudinally and transversely polarized beams has been given
and the observables sensitive to $CP$ violation in the $\T_i$
sector and Higgs sector have been classified.

On the other hand, also the $CP$ conserving observables depend
on the phases of the underlying complex
parameters, because the mass eigenvalues and the couplings
involved are functions of these parameters. In
particular, the various decay branching ratios depend in a
characteristic way on the complex
phases. The main purpose of the present paper
is a detailed study of the fermionic decay branching
ratios of $\ti \tau_1$, $\ti \tau_2$ and $\ti \nu_{\tau}$, and
the bosonic decay branching ratios of $\ti \tau_2$ and 
$\ti \nu_{\tau}$ and their dependences on the phases
$\varphi_{A_{\tau}}$, $\varphi_{\mu}$ and $\varphi_{U(1)}$. In
\cite{shortpaper} we have published first results of our study.
In the present paper we give the analytic expressions for the
various decay widths with complex couplings. We present a more 
detailed numerical study of the phase dependences of the various
branching ratios. We also discuss how these phase dependences
can be qualitatively 
understood on the basis of the analytic expressions for the
decay widths. Furthermore, we give a theoretical estimate of the
precision to be 
expected for the determination of the complex phases
together with the other fundamental parameters of the 
$\ti \tau_i$-system by measurements of suitable decay
branching ratios as well as masses and polarized
production cross sections in $e^+ e^-$ annihilation. 
Finally, we calculate the EDM of the
$\tau$-lepton induced by the $\tau$-slepton--neutralino and
$\tau$-sneutrino--chargino loops with complex $A_{\tau}$, $\mu$
and $M_1$.

In Section~2 we shortly review the 
mixing of 3rd generation sleptons in the presence of complex
parameters. 
In Section~3 we give the formulae for the fermionic and
bosonic decay widths of 
$\ti \tau_i$ and $\ti \nu_{\tau}$. In Section~4 we present 
numerical results for the phase dependences of their branching 
ratios. In Section~5 we give an estimate of
the errors to be expected for the fundamental parameters and the
phases of $A_{\tau}$, $\mu$ and $M_1$. In Section~6 we present
our results for the EDM of the $\tau$. Section~7 contains a
short summary.

\section{$\T_L - \T_R$ Mixing}

We first give a short account of $\T_L - \T_R$ mixing in the
case the parameters $\mu$ and 
$A_{\tau}$ are complex. The masses and couplings of the
$\tau$-sleptons follow from the 
hermitian $2 \times 2$ mass matrix which in the basis 
$(\ti \tau_L, \ti \tau_R)$ reads \cite{elru,guha}
\begin{equation}
{\mathcal{L}}_M^{\T}= -(\T_L^{\ast},\, \T_R^{\ast})
\left(\begin{array}{ccc}
M_{\T_{LL}}^2 & e^{-i\varphi_{\T}}|M_{\T_{LR}}^2|\\[5mm]
e^{i\varphi_{\T}}|M_{\T_{LR}}^2| & M_{\T_{RR}}^2
\end{array}\right)\left(
\begin{array}{ccc}
\T_L\\[5mm]
\T_R \end{array}\right),
\label{eq:mm}
\end{equation}
with
\begin{eqnarray}
M_{\T_{LL}}^2 & = & M_{\tilde L}^2+(-\frac{1}{2}+\sin^2\Theta_W)
\cos2\beta \ m_Z^2+m_{\tau}^2 ,\\[3mm]
M_{\T_{RR}}^2 & = & M_{\tilde E}^2-\sin^2\Theta_W\cos2\beta \
m_Z^2+m_{\tau}^2 ,\\[3mm]
 M_{\T_{RL}}^2 & = & (M_{\T_{LR}}^2)^{\ast}=
  m_{\tau}(A_{\tau}-\mu^{\ast}  
 \tan\beta), \label{eq:mlr}
\end{eqnarray}
\begin{equation}
\varphi_{\T}  = \arg\lbrack A_{\tau}-\mu^{\ast}\tan\beta\rbrack ,
\label{eq:phtau}
\end{equation}
where 
$m_{\tau}$ is the mass of the $\tau$-lepton, 
$\Theta_W$ is the weak mixing angle,
$\tan\beta=v_2/v_1$ with $v_1 (v_2)$ being the vacuum 
expectation value of the Higgs field $H_1^0 (H_2^0)$,
and $M_{\ti L}$, $M_{\ti E}, A_{\tau}$ are the soft
SUSY--breaking parameters of 
the $\T_i$ system. The $\T$ mass eigenstates are 
$(\tilde\tau_1, \tilde \tau_2)=(\T_L, \T_R)
{\mathcal{R}^{\T}}^T$ with 
 \begin{equation}
\mathcal{R}^{\T}=\left( \begin{array}{ccc}
e^{i\varphi_{\T}}\cos\theta_{\T} & 
\sin\theta_{\T}\\[5mm]
-\sin\theta_{\T} & 
e^{-i\varphi_{\T}}\cos\theta_{\T}
\end{array}\right),
\label{eq:rtau}
\end{equation}
and
\begin{equation}
\cos\theta_{\T}=\frac{-|M_{\T_{LR}}^2|}{\sqrt{|M_{\T _{LR}}^2|^2+
(m_{\T_1}^2-M_{\T_{LL}}^2)^2}},\quad
\sin\theta_{\T}=\frac{M_{\T_{LL}}^2-m_{\T_1}^2}
{\sqrt{|M_{\T_{LR}}^2|^2+(m_{\T_1}^2-M_{\T_{LL}}^2)^2}}.
\label{eq:thtau}
\end{equation}

The mass eigenvalues are
\begin{equation}
 m_{\T_{1,2}}^2 = \frac{1}{2}\left((M_{\T_{LL}}^2+M_{\T_{RR}}^2)\mp 
\sqrt{(M_{\T_{LL}}^2 - M_{\T_{RR}}^2)^2 +4|M_{\T_{LR}}^2|^2}\right).
\label{eq:m12}
\end{equation}

The $\snu_\tau$ appears only in the left--state. 
Its mass is given by
\beq
  m^2_{\ti\nu_\tau} = M^2_{\ti L} + \frac{1}{2} m^2_Z\cos 2\beta \,.
\label{eq:msnu}
\eeq

Eqs.~(\ref{eq:thtau}) and (\ref{eq:m12}) show that the phase
dependence of the mixing 
angle $\theta_{\T}$ and the eigenvalues $m_{\T_{1,2}}$ stems from
the term 
$m^2_\tau |A_{\tau}| |\mu| \tan\beta \cos (\varphi_{\mu} +
\varphi_{A_{\tau}})$. The phase dependence of $\theta_{\tilde \tau}$
is strongest if 
$|A_{\tau}| \approx |\mu|\tan\beta$ and  at the same time 
$|M_{\T_{LL}}^2 - M_{\T_{RR}}^2|\lsim|M_{\T_{LR}}^2|$.
The masses $m_{\tilde \tau_{1,2}}$ are in many cases insensitive to the phases 
$\varphi_{\mu}$ and $\varphi_{A_{\tau}}$ because $m_\tau$ is small.

\section{Production and Decay Formulae of $\T_i$ and 
$\ti \nu_{\tau}$} 

The reaction $e^+ e^- \ra \T_i \bar{\T_j}$ proceeds via
$\gamma$ and $Z$ exchange in the $s$-channel. The 
$Z \T_i \T_j$ couplings are 
\beqa
C(\T_1^{\ast} Z \T_1) & = &\frac{1}{2\cos\Theta_W} 
(\cos^2\theta_{\T} - 2\sin^2\Theta_W),\nonumber\\
C(\T_2^{\ast} Z \T_2) & = &\frac{1}{2\cos\Theta_W} 
(\sin^2\theta_{\T} - 2\sin^2\Theta_W),\nonumber\\
C(\T_2^{\ast} Z \T_1) & = &-\frac{1}{2\cos\Theta_W} e^{-i\varphi_{\T}}
\cos\theta_{\T}\sin\theta_{\T}\nonumber\\[2mm]
C(\T_1^{\ast} Z \T_2) & = & \lbrack C(\T_2^* Z \T_1)\rbrack^{\ast}.
\label{eq:zkop}
\eeqa
The reaction $e^+e^- \to \ti\nu_{\tau} \bar{\ti\nu_{\tau}}$
proceeds via $s$-channel $Z$ exchange with the coupling
\beq
C(\SN^{\ast} Z \SN) = - \frac{1}{2\cos\Theta_W} \, . 
\label{eq:zsnukop}
\eeq
The cross section of 
$e^+e^- \to \ti\nu_{\tau} \bar{\ti\nu_{\tau}}$ at tree level
does not depend 
on the phases $\varphi_{\mu}$ and $\varphi_{A_{\tau}}$.
The tree--level cross sections of the reactions 
$e^+ e^- \ra \T_i \bar{\T_j}$ do not explicitly depend on the
phases $\varphi_{\mu}$ and $\varphi_{A_{\tau}}$, because
the couplings $C(\T_i^{\ast} Z \T_i), i=1,2$, are real and in
$e^+ e^- \ra \T_1 \bar{\T_2}$ only $Z$ exchange
contributes. The cross sections depend only on the mass 
eigenvalues $m_{\T_{1,2}}$ and on the mixing angle
$\theta_{\T}$. Therefore, they depend only implicitly on 
the phases via 
the $\cos(\varphi_{\mu} + \varphi_{A_{\tau}})$ dependence of 
$m_ {\T_{1,2}}$ and $\theta_{\T}$, Eqs.~(\ref{eq:thtau}) and
(\ref{eq:m12}). This holds even if one or both beams are
polarized (the formulae of the cross sections including beam
polarizations are given, e. g., in \cite{sfermZP}). 
Of course, properly polarized $e^-$ and $e^+$ beams
are a very useful tool to 
enhance some signals and reduce the background and, therefore,
measure some of the observables with better precision
\cite{sferm,gust}. 
Information about the phases $\varphi_{\mu}$ and
$\varphi_{A_{\tau}}$ separately can be obtained by
studying the branching ratios of the $\T_i$ and $\SN$ decays
into neutralinos, charginos and Higgs bosons, because some of
them depend explicitly on the phases. 
It is expected that Yukawa--type
corrections at one--loop order to the $\ti \tau_i$ and 
$\ti \nu_{\tau}$ pair production cross sections and decay widths
will not change the overall picture obtained in tree
approximation, because they have been shown to be of the order
of a few percent only \cite{yukcorr}.

\subsection{Fermionic Decay Widths of $\T_i$ and
$\ti\nu_{\tau}$} 

The widths for the decays
$\T_i\rightarrow \CH_j^0 \tau (\lambda_{\tau}), i=1,2,
j=1,\dots, 4$, 
where $\CH_j^0$ is the neutralino and 
$\lambda_{\tau} = \pm \frac{1}{2}$
is the helicity of the outgoing $\tau$, read
\begin{equation}
\Gamma(\T_i\rightarrow \CH_j^0 \tau (\lambda_{\tau})) = 
\frac{g^2 \kappa(m_{\T_i}^2,m_{\CH_j^0}^2,m_{\tau}^2)}
{16\pi m_{\T_i}^3}\ |\mathcal{M}_{\lambda_{\tau}}|^2\quad 
\label{eq:widthferm}
\end{equation} 
with
\begin{eqnarray}\label{amplitude}
 |\mathcal{M}_{\lambda_{\tau}}|^2 &=& \frac{1}{4}
 \biggl(H_s^2\Bigl(|b_{ij}^{\T}|^2+|a_{ij}^{\T}|^2+2\Re e(
 {b_{ij}^{\T}}^{\ast}a_{ij}^{\T})\Bigr)+{}\nonumber\\[5mm]
 &&{}+H_p^2\Bigl(|b_{ij}^{\T}|^2+|a_{ij}^{\T}|^2-2\Re
 e({b_{ij}^{\T}}^{\ast}a_{ij}^{\T})\Bigr)+{}\nonumber\\[5mm]
 &&{}+2\,(-1)^{\lambda_{\tau}+\frac{1}{2}}\,
 H_pH_s\Bigl(|a_{ij}^{\T}|^2-|b_{ij}^{\T}|^2\Bigr)
 \Biggr)\quad 
 \end{eqnarray}
where $g$ is the weak $SU(2)$ gauge coupling constant, 
$\kappa(x,y,z)=(x^2+y^2+z^2-2xy-2xz-2yz)^{1/2}$ and
$ H_s=(m_{\T_i}^2-(m_{\CH_j^0}+m_{\tau})^2)^{\frac{1}{2}},
H_p=(m_{\T_i}^2-(m_{\CH_j^0}- m_{\tau})^2)^{\frac{1}{2}}$.
The couplings are
\begin{equation}
a_{ij}^{\T}=(\mathcal{R}^{\T}_{in})^{\ast}\mathcal{A}^{\tau}_{jn},\qquad 
b_{ij}^{\T}=(\mathcal{R}^{\T}_{in})^{\ast}\mathcal{B}^{\tau}_{jn},\qquad
\ell_{ij}^{\T}=(\mathcal{R}^{\T}_{in})^{\ast}\mathcal{O}^{\tau}_{jn}\qquad
(n=L,R)
\label{eq:coupl1}
\end{equation}
where
\begin{equation}
\mathcal{A}^{\tau}_j=\left(\begin{array}{ccc}
\f_{Lj}\\[2mm]
\h_{Rj} \end{array}\right),\qquad \mathcal{B}^{\tau}_j=\left(\begin{array}{ccc}
\h_{Lj}\\[2mm]
\f_{Rj} \end{array}\right),
\qquad \mathcal{O}^{\tau}_j=\left(\begin{array}{ccc}
- U_{j1}\\[2mm]
Y_{\tau} U_{j2} \end{array}\right)
\label{eq:coupl2}
\end{equation}
with
\begin{eqnarray}
\h_{Lj}&=& (\h_{Rj})^{\ast}=Y_{\tau} N_{j3}^{\ast}\nonumber \\  
\f_{Lj}&=& -\frac{1}{\sqrt{2}}(\tan\Theta_W N_{j1}+N_{j2})\nonumber\\
\f_{Rj}&=& \sqrt{2}\tan\Theta_W N_{j1}^{\ast}.
\label{eq:coupl3}
\end{eqnarray}
$Y_{\tau}= m_{\tau}/(\sqrt{2}m_W \cos\beta)$ is the $\tau$
Yukawa coupling. The mixing
matrices $U$ and $N$ are defined by Eqs.~(\ref{eq:Uchar}) and 
(\ref{eq:mixN}) in Appendices~\ref{app:char} and \ref{app:neut}.
Since $m_{\tau}\ll m_{\T_1}$, we have $H_s \approx H_p$ and,
hence, to a good approximation, 
$\Gamma(\T_i\to \CH_j^0 \tau (\lambda_{\tau})) \propto
|b_{ij}^{\T}|^2 
(|a_{ij}^{\T}|^2)$ for $\lambda_{\tau}=+\frac{1}{2} 
(-\frac{1}{2})$ \cite{nojiri}.

The width for the decay into the chargino,
$\T_i \rightarrow \CH_j^- \nu_{\tau} (i,j=1,2)$, 
is obtained by the replacements 
$a_{ij}^{\T}\to \ell_{ij}^{\T}, 
b_{ij}^{\T}\to 0, m_{{\CH}_j^0}\to m_{{\CH}_j^-}, 
m_{\tau}\to 0$ and $\lambda_{\tau}\to -\frac{1}{2}$
in Eqs.~(\ref{eq:widthferm}) and 
(\ref{amplitude}), with the couplings $\ell_{ij}^{\T}$ 
also given in Eqs.~(\ref{eq:coupl1}) and (\ref{eq:coupl2}). 
The width for the $\tau$-sneutrino decay
$\SN \rightarrow \CH_j^0\nu_{\tau}$ is obtained by the
replacements $a_{ij}^{\T} \to a_j^{\tilde\nu}$,
$b_{ij}^{\T} \to 0$, $m_{{\tilde \tau}_i} \to m_{{\tilde \nu}_\tau}$,
$m_\tau \to 0$ and 
$\lambda_{\tau} \to -\frac{1}{2}$
in Eqs.~(\ref{eq:widthferm}) and (\ref{amplitude}), and that 
for the decay $\SN \rightarrow \CH_j^+\tau (\lambda_{\tau})$ by
the replacements $a_{ij}^{\T}\to \ell_j^{\tilde\nu}, 
b_{ij}^{\T}\to k_j^{\tilde\nu}$,
$m_{{\tilde \tau}_i} \to m_{{\tilde \nu}_\tau}$ and
$m_{{\tilde \chi}^0_j} \to m_{{\tilde \chi}^+_j}$. The couplings are now 
\begin{equation}
a^{\tilde\nu}_j=\frac{1}{\sqrt{2}}(N_{j1}\tan\Theta_W - N_{j2})
,\qquad
 k^{\tilde\nu}_j=
 Y_{\tau} U_{j2}^{\ast},\qquad\ell^{\tilde\nu}_j=-V_{j1} \, ,
\label{eq:coupl4}
\end{equation}
with the mixing matrix $V$ given by
 Eq.~(\ref{eq:Vchar}) in Appendix~\ref{app:char}.

As can be seen, the widths for the decays of $\T_1$ and $\T_2$ 
into charginos and neutralinos depend on 
$\cos(\varphi_{\mu} + \varphi_{A_{\tau}})$ 
through $m_{\T_i}$ and $\theta_{\tilde \tau}$, and also on 
$\varphi_{\T}$, Eq.~(\ref{eq:phtau}). They depend also on
$\varphi_{\mu}$ ($\varphi_{\mu}$ and $\varphi_{U(1)}$) via the
chargino (neutralino) masses $m_{\CH_j^-} (m_{\CH_j^0})$ 
and mixing matrix $U (N)$, see
Eqs.~(\ref{eq:massCH}-\ref{eq:massch}) 
(Eqs.~(\ref{eq:massN},\ref{eq:mixN})).
The widths for the $\tilde \nu_\tau$ decays into fermions 
depend on the phases of
$\varphi_{\mu}$ and $\varphi_{U(1)}$.

\subsection{Bosonic Decay Widths of $\T_2$ and
$\ti\nu_{\tau}$} 

The widths for the decays of $\T_2$ and $\ti \nu_{\tau}$ into
gauge bosons and Higgs bosons are given by: 
\begin{equation} 
\Gamma(\T_2 \rightarrow W^- \SN) =
\frac{g^2\kappa^3 (m_{\T_2}^2,m_{\SN}^2,m_{W^{\pm}}^2)}{16\pi
m_{\T_2}^3 m_{W^{\pm}}^2}\ |C(\SN^{\ast} W^+ \T_2)|^2
\label{eq:Wwidth}
\end{equation}
\begin{equation} 
\Gamma(\T_2 \rightarrow Z \T_1) =
\frac{g^2\kappa^3(m_{\T_2}^2,m_{\T_1}^2,m_{Z}^2)}{16\pi
m_{\T_2}^3 m_{Z}^2}\ |C(\T^{\ast}_1 Z \T_2)|^2
\label{eq:Zwidth}
\end{equation}
\begin{equation} 
\Gamma(\T_2 \rightarrow H^- \SN) =
\frac{g^2\kappa(m_{\T_2}^2,m_{\SN}^2,m_{H^{\pm}}^2)}{16\pi
m_{\T_2}^3}\ |C(\SN^{\ast} H^+ \T_2)|^2
\label{eq:chHwidth}
\end{equation}
\begin{equation} 
\Gamma(\T_2 \rightarrow H_i \T_1) =
\frac{g^2\kappa(m_{\T_2}^2,m_{\T_1}^2,m_{H_i}^2)}{16\pi
m_{\T_2}^3}\ |C(\T^{\ast}_1 H_i \T_2)|^2
\label{eq:H0width}
\end{equation}
\begin{equation} 
\Gamma(\SN \rightarrow W^+ \T_1) =
\frac{g^2\kappa^3 (m_{\SN}^2,m_{\T_1}^2,m_{W^{\pm}}^2)}{16\pi
m_{\SN}^3 m_{W^{\pm}}^2}\ |C(\T_1^{\ast} W^- \SN)|^2
\label{eq:snWwidth}
\end{equation}
\begin{equation} 
\Gamma(\SN \rightarrow H^+ \T_1) =
\frac{g^2\kappa(m_{\SN}^2,m_{\T_1}^2,m_{H^{\pm}}^2)}{16\pi
m_{\SN}^3}\ |C(\T_1^{\ast} H^- \SN)|^2
\label{eq:snchHwidth}
\end{equation}
 
The couplings relevant for $\T_2$ decays into the $Z$ boson 
are given in Eq. (\ref{eq:zkop}) and the couplings to the $W^+$
boson are 
\begin{equation}
C(\SN^{\ast} W^+\T_{1,2})=\frac{1}{\sqrt{2}}
(- e^{-i\varphi_{\ti \tau}} \cos\theta_{\T}, \sin\theta_{\T}).
\label{eq:wkop}
\end{equation}

The couplings to the Higgs bosons are more conveniently written
in the weak basis $(\T_L, \T_R)$. The couplings to the charged
Higgs boson $H^+$ are given by 
\begin{equation}\label{chHcoup}
C(\SN^{\ast} H^+ \T_{L,R})=\frac{1}{\sqrt{2} m_W}
\left(m_{\tau}^2 \tan\beta-m_W^2
\sin2\beta,m_{\tau}(\tan\beta |A_{\tau}| e^{-i\varphi_{A_{\tau}}}+
|\mu|e^{i\varphi_{\mu}})\right)
\end{equation}
\\
The couplings $C(\SN^{\ast} H^+ \T_{1,2})$ of the mass
eigenstates $\T_i$ are then obtained
by multiplying the couplings above with 
${\mathcal{R}^{\T}}^{\dagger}$ from the right.

The couplings to the neutral Higgs bosons $H_i, i=1,2,3,$ are 
\begin{equation}
C(\T_{L}^{\ast} H_i \T_{L})=-\frac{m_{\tau}^2}{m_W
\cos\beta}\mathcal{O}_{1i} 
-\frac{m_Z}{\cos\Theta_W}\left(-\frac{1}{2}+\sin^2\Theta_W\right)
\left(\cos\beta\mathcal{O}_{1i} 
-\sin\beta\mathcal{O}_{2i}\right)
\label{eq:HLLcoup}
\end{equation}

\begin{equation}
C(\T_{R}^{\ast} H_i \T_{R})=-\frac{m_{\tau}^2}{m_W
\cos\beta}\mathcal{O}_{1i} 
+\frac{m_Z}{\cos\Theta_W} \sin^2\Theta_W(\cos\beta \mathcal{O}_{1i}
-\sin\beta\mathcal{O}_{2i}),
\label{eq:HRRcoup}
\end{equation}

\begin{eqnarray}
C(\T_{L}^{\ast} H_i \T_{R})&=& \frac{m_{\tau}}{2 m_W
\cos\beta}\{i \left(\sin\beta |A_{\tau}| e^{-i\varphi_{A_{\tau}}}
+\cos\beta |\mu|e^{i\varphi_{\mu}}\right) \mathcal{O}_{3i}
\nonumber\\[3mm]
&&{}+\left(|\mu|e^{i\varphi_{\mu}}\mathcal{O}_{2i}- 
|A_{\tau}| e^{-i\varphi_{A_{\tau}}} \mathcal{O}_{1i}\right)\},
\label{eq:HLRcoup}
\end{eqnarray}

\begin{equation}
C(\T_{R}^{\ast} H_i \T_{L})=\lbrack C(\T_{L}^{\ast} H_i \T_{R})
\rbrack^{\ast}.
\end{equation}
\\
The couplings of the  mass eigenstates $\T_i$ are obtained by 
\begin{equation}
C(\T_k^{\ast} H_i \T_j)= \mathcal{R}^{\T}\cdot
\left(\begin{array}{ccc}
C(\T_{L}^{\ast} H_i \T_{L}) & C(\T_{L}^{\ast} H_i \T_{R})\\[5mm]
 C(\T_{R}^{\ast} H_i \T_{L}) & C(\T_{R}^{\ast} H_i \T_{R})
\end{array}\right)\cdot {\mathcal{R}^{\T}}^{\dag} .
\label{eq:Hijcoup}
\end{equation}
\\
$\mathcal{O}_{ij}$ is the real orthogonal mixing matrix in the
neutral Higgs sector in the basis 
$(\phi_1, \phi_2, a)=
(\sqrt{2} (\mathcal{R}e H^0_1 -v_1)$ ,
$\sqrt{2} (\mathcal{R}e H^0_2 -v_2)$,
$\sqrt{2} (\sin\beta \mathcal{I}m H^0_1 + 
\cos\beta \mathcal{I}m H^0_2))$, where $H_1^0$ and $H_2^0$ are
the neutral members of the two Higgs doublets with hypercharge
$-1$ and $+1$, respectively.
$\mathcal{O}_{ij}$ diagonalises the $3\times 3$
Higgs mass matrix: $\phi_i = \mathcal{O}_{ij} H_j, i=1,2$, 
$a = \mathcal{O}_{3j} H_j$, 
$\mathcal{O}^{T} \mathcal{M}_{H}^2 \mathcal{O}=
diag(m_{H_1}^2,m_{H_2}^2,m_{H_3}^2)$, 
with $m_{H_1}\leq m_{H_2}\leq m_{H_3}$ \cite{carenaetal}. The 
neutral Higgs mass eigenstates $H_i, i= 1,2,3$, are mixtures of
the $CP$-even and $CP$-odd states, because of the explicit $CP$ 
violation in the Higgs sector. The phase parameter
$\xi$ also introduced in \cite{carenaetal,pilaftsi1,dreescp}
does not play a role in our analysis. Therefore we put 
$\xi = 0$. 

The widths for $\T_2$ decays into the neutral
Higgs bosons depend on 
$\varphi_{\mu}$, $\varphi_{A_{\tau}}$ and $\varphi_{\T}$
and in addition on the mixing matrix 
$\mathcal{O}_{ij}$. At one--loop level
$\mathcal{O}_{ij}$ depends on the phases
$\varphi_{\mu}$, $\varphi_{A_t}$ and $\varphi_{A_b}$, with the
latter two being the phases of the stop and the sbottom trilinear
couplings $A_t$ and $A_b$, respectively. 

\section{Numerical Results}

In the following we present our numerical results showing
how the $\T_1$, $\T_2$ and $\tilde \nu_{\tau}$ decay branching
ratios depend on the complex phases. In order to study the full
phase dependences of the observables, we do not take into
account the restrictions on $\varphi_{\mu}$ and $\varphi_{U(1)}$
from the electron and 
neutron EDMs. We fix the $\T_1$, $\T_2$
and $\tilde \nu_{\tau}$ masses such that these particles can be pair produced
at an $e^+e^-$ linear collider with a CMS energy in the range
$\sqrt{s}=0.5 - 1.2$~TeV. Furthermore, we impose the following
conditions: 
\begin{itemize}
\item[(i)] $m_{\CH_1^{\pm}}>103$~GeV, $m_{H_1}>110$~GeV, 
$m_{\T_1}> m_{\CH_1^0}> 50$~GeV, $m_{\T_1} > 80$~GeV, and 
\item[(ii)] $|A_{\tau}|^2 < 3(M_{\ti L}^2+M_{\ti E}^2+
   (m_{H^+}^2+m_Z^2 \sin^2 \Theta_W )\sin^2\beta-\frac{1}{2}m_Z^2)$ (the
approximate necessary condition for tree--level vacuum 
stability \cite{casas}). 
\end{itemize}

In principle, the experimental data for the rare decay 
$b \to s \gamma$ lead to strong constraints on the SUSY and
Higgs parameters in the MSSM and, in particular, in the minimal
Supergravity Model (mSUGRA). 
We do not impose this constraint, because it strongly
depends on the detailed properties of the squarks, in particular
on the mixing between the squark families, which we do not take
into account.

The following parameters are necessary to specify the masses and
couplings of the SUSY particles $\T_i$, $\SN$, 
$\ti \chi_i^{\pm}$ and $\ti \chi_j^0$: $M_{\ti L}$, $M_{\ti E}$,
$|A_{\tau}|$, $\varphi_{A_{\tau}}$, $|\mu|$, $\varphi_{\mu}$,
$\tan\beta$, $M_2$, $|M_1|$, $\varphi_{U(1)}$. Equivalently  
we use the mass eigenvalues $m_{\T_1}$, $m_{\T_2}$ or the
masses $m_{\T_1}$, $m_{\SN}$ as input parameters instead of
$M_{\ti L}$, $M_{\ti E}$. 
For the complete determination of the renormalization group (RG)
improved MSSM Higgs sector at
one--loop level in addition the charged Higgs boson mass 
$m_{H^{\pm}}$, the mass parameters and the trilinear couplings
of the scalar top and scalar bottom systems 
$M_{\ti Q}$, $M_{\ti U}$, $M_{\ti D}$,
$|A_t|$, $\varphi_{A_t}$, $|A_b|$, $\varphi_{A_b}$ and the gluino
mass $|m_{\tilde g}|$ 
as well as its phase $\varphi_{\tilde g} = \arg(m_{\tilde g})$ have to be
specified \cite{carenaetal}. Mixing of the $CP$-even and
$CP$-odd neutral Higgs bosons at one--loop level is induced if
$A_{b,t}$ and/or $\mu$ are complex. 
We take  $m_\tau = 1.78$ GeV, $m_t=175$ GeV, $m_b=5$ GeV, 
$m_Z=91.2$ GeV, $\sin^2\Theta_W =0.23$, $m_W = m_Z \cos\Theta_W$, 
$\a(m_Z)=1/129$, and $\alpha_s(m_Z)=0.12$, 
where $m_{t,b}$ are pole masses of t and b quarks.

\subsection{$\T_1$ Decays}
In this subsection we study the dependence of the
branching ratios of $\T_1$ decays into charginos and
neutralinos on the phases $\varphi_{A_{\tau}}$, 
$\varphi_{\mu}$ and $\varphi_{U(1)}$. We take
$m_{\T_1}=240$~GeV. In order not to vary too many parameters we
fix $|A_{\tau}|=1000$~GeV in Figs.~1 to 7. We assume the GUT relation
$|M_1|=(5/3) \tan^2\Theta_W M_2$, although we take
$M_1$ complex. We focus on the decays 
$\T_1\rightarrow \CH^0_{1,2} \tau$ and
$\T_1\rightarrow \CH^-_1 \nu_{\tau}$.

\begin{figure}[H]
\begin{center}
\includegraphics{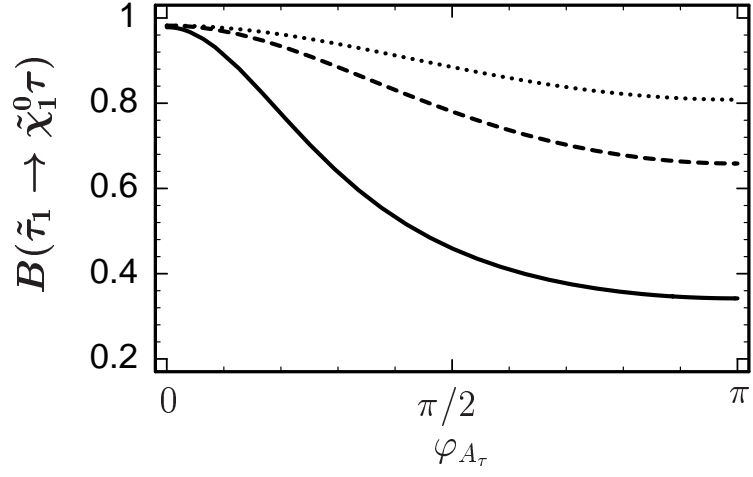}
\end{center}
\caption{Branching ratio of $\T_1 \to \CH_1^0 \tau$ as a
function of $\varphi_{A_{\tau}}$ for 
$m_{\T_1}=240$~GeV, $m_{\SN}=233$~GeV~$(\rm solid~line)$, 
$238$~GeV~$(\rm dashed~line)$, $243$~GeV~$(\rm dotted~line)$,
and $\varphi_{\mu} = \varphi_{U(1)} = 0$,
$|\mu| = 300$~GeV, $|A_{\tau}| = 1000$~GeV, $\tan\beta = 3$, and
$M_2 = 200$~GeV.} 
\label{fig:stau1.1}
\end{figure}

We first study the $\varphi_{A_{\tau}}$
dependence of the $\T_1$ decay branching ratios, because 
$\varphi_{A_{\tau}}$ appears only in the $\T_i$ sector and it is
the phase dependence that we are particularly interested in. In
Fig.~\ref{fig:stau1.1} we plot 
the branching ratio $B(\T_1 \ra \CH^0_1\tau)$ as a function of 
$\varphi_{A_{\tau}}$ 
for the three values $m_{\SN} = 233$~GeV, $238$~GeV and
$243$~GeV (corresponding to  
$M_{\ti L}=240$~GeV, $245$~GeV and $250$~GeV), taking 
$\varphi_{\mu} = \varphi_{U(1)} = 0$,
$|\mu| = 300$~GeV, $\tan\beta = 3$, and $M_2 = 200$~GeV.
Note that $B(\T_1 \ra \CH^0_1\tau)$ is invariant under
$\varphi_{A_{\tau}} \to -\varphi_{A_{\tau}}$ for
$\varphi_{\mu}=\{0,\pm \pi\}$ and 
$\varphi_{U(1)}= \{0,\pm \pi\}$.
As can be seen, the $\varphi_{A_{\tau}}$ dependence of 
$B(\T_1 \ra \CH^0_1\tau)$ is quite pronounced. To a large 
extend it is caused by a relatively strong variation of the mixing
angle $\theta_{\T}$ with varying $\varphi_{A_{\tau}}$.
More specifically, when varying $\varphi_{A_{\tau}}$ from
$0$ to $\pi$, then $\cos\theta_{\T}$ varies from $-0.1$ to
$-0.9$ for $m_{\SN} = 233$~GeV, from $-0.06$ to $-0.6$
for $m_{\SN} = 238$~GeV and from $-0.05$ to $-0.45$ 
for $m_{\SN} = 243$~GeV. This means that for 
$m_{\SN} = 238$~GeV and $243$~GeV $\T_1$ is mainly
$\T_R$-like, whereas for $m_{\SN} = 233$~GeV $\T_1$ is
$\T_L$-like ($\T_R$-like) for 
$\varphi_{A_{\tau}} \gsim \pi/3 (\lsim \pi/3)$.
Such a strong variation of the mixing angle $\theta_{\T}$ with
$\varphi_{A_{\tau}}$ can only occur if 
$M_{\ti L} \approx M_{\ti E}$ {\it and} 
$|A_{\tau}| \approx |\mu| \tan\beta$, otherwise this variation
is weaker.

\begin{figure}[H]
\begin{center}
\includegraphics{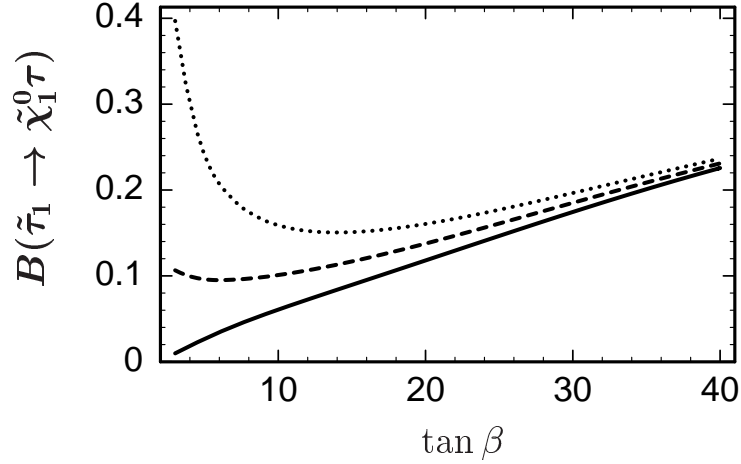}
\end{center}
\caption{Branching ratio of $\T_1 \to \CH_1^0 \tau$ as a
function of $\tan\beta$ for 
$m_{\T_1}=240$~GeV, $m_{\T_2}=500$~GeV,
$\varphi_{\mu} = 0 (\rm solid~line)$, 
$\pi/2 (\rm dashed~line)$, $\pi (\rm dotted~line)$, with 
the other parameters $\varphi_{A_{\tau}}=  
\varphi_{U(1)}=0$, $M_2=200$~GeV, $|\mu|=150$~GeV, and
$|A_{\tau}| = 1000$~GeV, assuming $M_{\ti L} < M_{\ti E}$.} 
\label{fig:stau1.2}
\end{figure}

In the following Figs.~2 to 5 we fix 
$m_{\T_2}=500$~GeV instead of $m_{\SN}$. We consider
separately the two cases 
$M_{\ti L} < M_{\ti E}$ and $M_{\ti L} \geq M_{\ti E}$ and
determine the values of $M_{\ti L}$ and $M_{\ti E}$
correspondingly. 
In Fig.~\ref{fig:stau1.2} we show the $\tan\beta$ dependence of 
$B(\T_1 \ra \CH^0_1\tau)$ for  
$\varphi_{\mu}=0$ (solid line), $\varphi_{\mu}=\pi/2$ (dashed line),
$\varphi_{\mu}=\pi$ (dotted line), with
$\varphi_{A_{\tau}} = \varphi_{U(1)} = 0$, $M_2=200$~GeV,
$|\mu|=150$~GeV, 
assuming $M_{\ti L} < M_{\ti E}$. For 
$\varphi_{A_{\tau}}=\{0,\pm \pi\}$ the branching ratios are
invariant under the simultaneous sign flip
$(\varphi_{\mu},\varphi_{U(1)})\to
(-\varphi_{\mu},-\varphi_{U(1)})$. As can be 
seen, $B(\T_1 \ra \CH^0_1\tau)$ becomes almost independent of
$\varphi_{\mu}$ for $\tan\beta \grts 15$. A similar behaviour is
obtained for $B(\T_1 \ra \CH^0_2\tau)$ and 
$B(\T_1 \ra \CH^-_1\nu_{\tau})$. In the case of the decay 
$\T_1 \ra \CH^-_1\nu_{\tau}$ this behaviour can be understood by
observing that the $\varphi_{\mu}$ dependence of the mass
eigenvalues $m_{\CH^{\pm}_i}$ and the mixing
matrices $U_{ij}$ and $V_{ij}$ changes if the value of
$\tan\beta$ is changed. For the width
$\Gamma(\T_1 \ra \CH^-_1\nu_{\tau}) \propto 
|\ell^{\ti \tau}_{11}|^2$ we obtain from Eqs.~(\ref{eq:rtau}),
 (\ref{eq:coupl1}),
(\ref{eq:coupl2}) and (\ref{eq:Uchar}) 
\beq
\nonumber
\ell^{\T}_{11} = -e^{i\gamma_1}
(e^{-i\varphi_{\ti \tau}}\cos\theta_{\T} \cos\theta_1 - 
e^{i\phi_1} Y_{\tau} \sin\theta_{\ti \tau} \sin\theta_1).
\label{eq:lapprox}
\eeq
By inspecting Eqs.~(\ref{eq:phtau}) and (\ref{eq:phi1}) one can 
verify that in the limit $\tan\beta \to \infty$ we obtain 
$e^{-i\varphi_{\ti \tau}} \to -e^{i\varphi_{\mu}}$ and
$e^{i\phi_1} \to e^{i\varphi_{\mu}}$, which means that in this
limit $|\ell^{\ti \tau}_{11}|$ becomes independent of 
$\varphi_{\mu}$. 
Here note that in this limit $\theta_{\tilde \tau}$ and $\theta_1$
become independent of $\varphi_\mu$ as can be seen from 
Eqs.~(\ref{eq:mlr}), (\ref{eq:thtau}), (\ref{eq:m12}) and (\ref{eq:theta1}).
In the case of the decay into a
neutralino we can see the influence of the phases
$\varphi_{\mu}$ and $\varphi_{U(1)}$ from the approximate
formulae 
\begin{equation}
m_{\CH_1^0}\simeq |M_1| \left(1-
\frac{m_Z^2 \sin^2\Theta_W \sin2\beta \cos(\varphi_{\mu}+
\varphi_{U(1)})}{|\mu| |M_1|}\right)
\label{eq:mapprox1}
\end{equation}  
and
\begin{equation}
m_{\CH_1^0}\simeq |\mu|\left(1-
\frac{m_Z^2}{2|\mu|}\Bigg\lbrace\left[\frac{\sin^2\Theta_W}{|M_1|}
+\frac{\cos^2\Theta_W}{M_2}\right]+\sin2\beta
\left[\frac{\sin^2\Theta_W 
\cos(\varphi_{\mu}+\varphi_{U(1)})}{|M_1|}+\frac{\cos^2\Theta_W
\cos\varphi_{\mu}}{M_2}\right]\Bigg\rbrace\right),
\label{eq:mapprox2}
\end{equation} 
which hold for $|M_2 \pm |\mu|| \gg m_Z$ for the mass of a
gaugino--like or a higgsino--like $\chi^0_1$, respectively.
Similar approximation formulae hold for $m_{\CH^0_2}$ 
and the mixing matrix $N_{ij}$.
From these formulae one can see
that $\varphi_{\mu}$ and $\varphi_{U(1)}$ appear only in terms
multiplied by $\sin2\beta$. Therefore, in the approximation
where Eqs.~(\ref{eq:mapprox1}) and (\ref{eq:mapprox2}) hold,
$m_{\CH^0_{1,2}}$ and $N_{ij}$ become independent of
$\varphi_{\mu}$ and $\varphi_{U(1)}$ for large $\tan\beta$.
Concerning the $\varphi_{\mu}$ dependence in general, it can be
shown that $m_{\CH^0_i}$ and $N_{ij}$ become independent 
of $\varphi_{\mu}$ for $\tan\beta \to \infty$, because the
characteristic equation of the neutralino mass eigenvalues
becomes independent of $\varphi_{\mu}$ in this limit. 

\begin{figure}[H]
\setlength{\unitlength}{1mm}
\begin{center}
\begin{picture}(150,60)
\put(-5,0){\mbox{\epsfig{figure=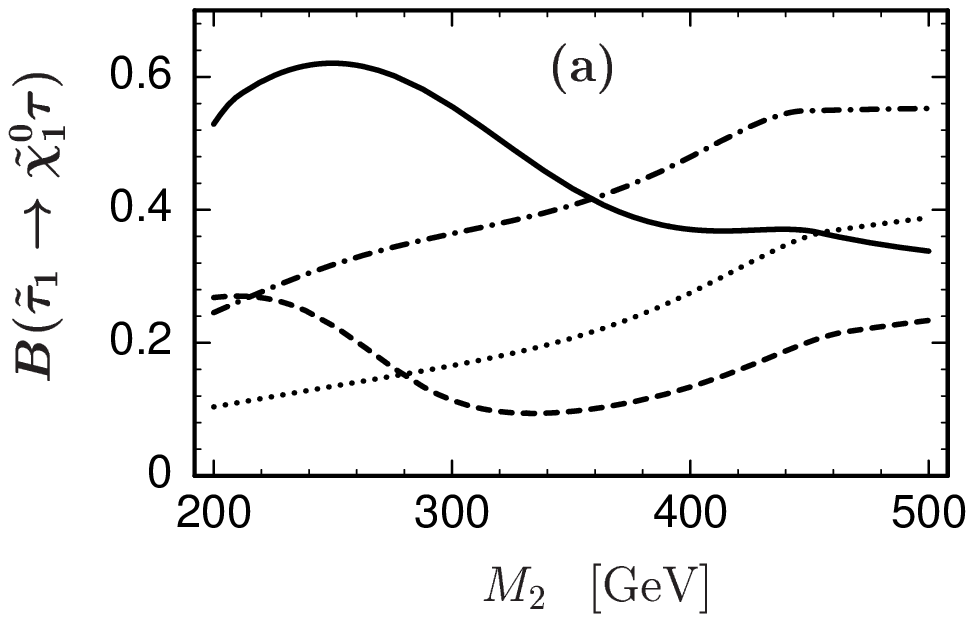,height=6.cm,width=8.cm}}}
\put(75,0){\mbox{\epsfig{figure=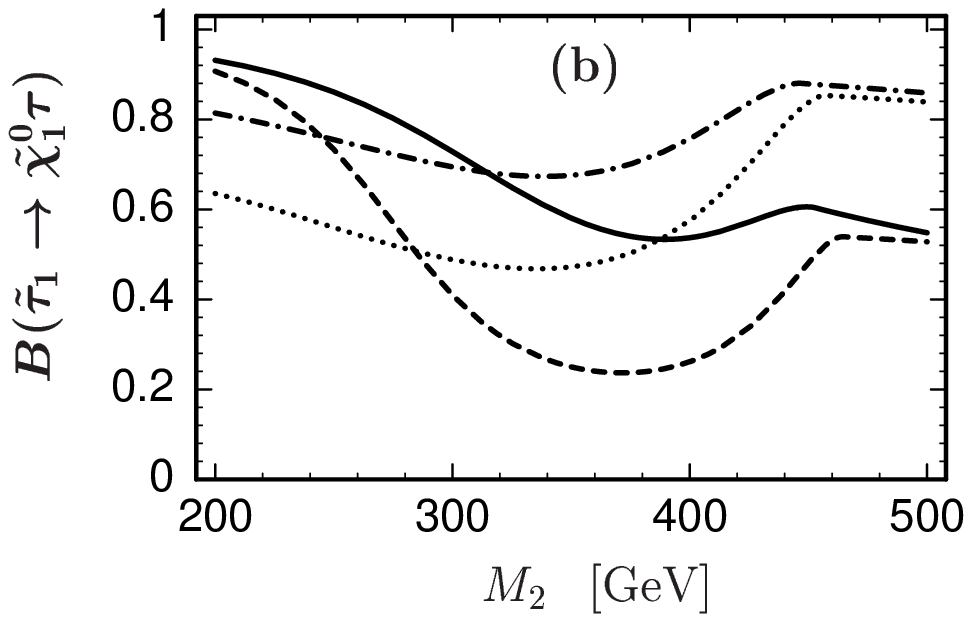,height=6.cm,width=8.cm}}}
\end{picture}
\end{center}
\caption{Branching ratio of $\T_1 \to \CH_1^0 \tau$ as a
function of $M_2$ for 
$\varphi_{\mu}=\pi (\rm solid~line)$, 
$\pi/2 (\rm dashed~line)$, $0 (\rm dotted~line)$, 
$-\pi/2 (\rm dashdotted~line)$, 
$\varphi_{A_{\tau}}=0$, $\varphi_{U(1)}=\pi/2$, 
$m_{\T_1}=240$~GeV, $m_{\T_2}=500$~GeV,
$|\mu|= 150$~GeV, $\tan\beta = 3$, and $|A_{\tau}| = 1000$~GeV, 
assuming a) $M_{\ti L} < M_{\ti E}$, 
b) $M_{\ti L} \geq M_{\ti E}$.}
\label{fig:stau1.3}
\end{figure}

In Figs.~\ref{fig:stau1.3}\,a,\,b we plot the
branching ratio $B(\T_1 \ra \CH^0_1\tau)$ against $M_2$
in the range $200$~GeV $\leq M_2 \leq 500$~GeV for 
$\varphi_{\mu}=\pi$ (solid line), $\varphi_{\mu}=\pi/2$ (dashed line),
$\varphi_{\mu}=0$ (dotted line) and $\varphi_{\mu}=-\pi/2$
(dash-dotted line),
taking $\varphi_{A_{\tau}}=0$, $\varphi_{U(1)}=\pi/2$, $|\mu|=
150$~GeV and $\tan\beta = 3$. In Fig.~\ref{fig:stau1.3}\,a we
assume $M_{\tilde L}< M_{\tilde E}$, so that  
$\T_1\simeq\T_L (\cos\theta_{\T}\approx
-1)$. This means that the couplings are approximately 
$|a_{1j}^{\T}|\simeq |\f_{Lj}|$, $|b_{1j}^{\T}|\simeq |\h_{Lj}|$
and the decay width is essentially determined by 
$\Gamma(\T_1 \ra \CH^0_j\tau) \propto |\f_{Lj}|^2 +
|\h_{Lj}|^2$. 
In Fig.~\ref{fig:stau1.3}\,b we
consider the case $M_{\tilde L}\geq M_{\tilde E}$.
In this case we have $|a_{1j}^{\T}|\simeq |\h_{Rj}|$, 
$|b_{1j}^{\T}|\simeq |\f_{Rj}|$ and 
$|\ell_{1j}^{\T}| \simeq Y_{\tau}|U_{j2}|$.
This means that the decay $\T_1 \ra \CH_1^-\nu_{\tau}$ is 
suppressed, because now $\T_1 \simeq \T_R$ 
($\cos\theta_{\T}\approx 0$) and the 
$\T_1 \CH_1^- \nu_{\tau}$ coupling is nearly proportional to the 
small Yukawa coupling $Y_{\tau}$. Therefore, $B(\T_1 \ra \CH^0_1\tau)$
in Fig.~\ref{fig:stau1.3}\,b is larger 
than in Fig.~\ref{fig:stau1.3}\,a. In both cases there
is a significant variation with $\varphi_{\mu}$. 
The $\varphi_{\mu}$ dependence of
$B(\T_1 \ra \CH^0_1\tau)$
in Figs.~\ref{fig:stau1.3}\,a,\,b is caused by an interplay
between the $\varphi_{\mu}$ dependence of the mass and mixing
character of the $\T_1$ and that of the $\tilde \chi^0_1$. The $M_2$
dependence can be understood by noting that for $M_2 \approx
200$~GeV the lightest neutralino has a sizable gaugino content,
which decreases for increasing $M_2$. For our parameter choice
$\ti \chi^0_1$ becomes mainly higgsino--like for $M_2 \grts
300$~GeV. Near $M_2 \approx 440$~GeV the decays into
gaugino--like neutralinos become kinematically forbidden, which causes
the increase of $B(\T_1 \ra \CH^0_1\tau)$ for 
$M_2 \grts 400$~GeV.

We have studied the $\varphi_{\mu}$ dependence of
$B(\T_1 \ra \CH^0_1\tau)$ also for other values of $|\mu|$ and
have found that it is less pronounced if $|\mu| \grts M_2$ and
that it is stronger if $|\mu| \approx M_2$ or 
$|\mu| \lets |M_1|$. As shown in Fig. \ref{fig:stau1.2} it is
stronger for low $\tan\beta$. 

\begin{figure}[H]
\setlength{\unitlength}{1mm}
\begin{center}
\begin{picture}(150,60)
\put(-5,0){\mbox{\epsfig{figure=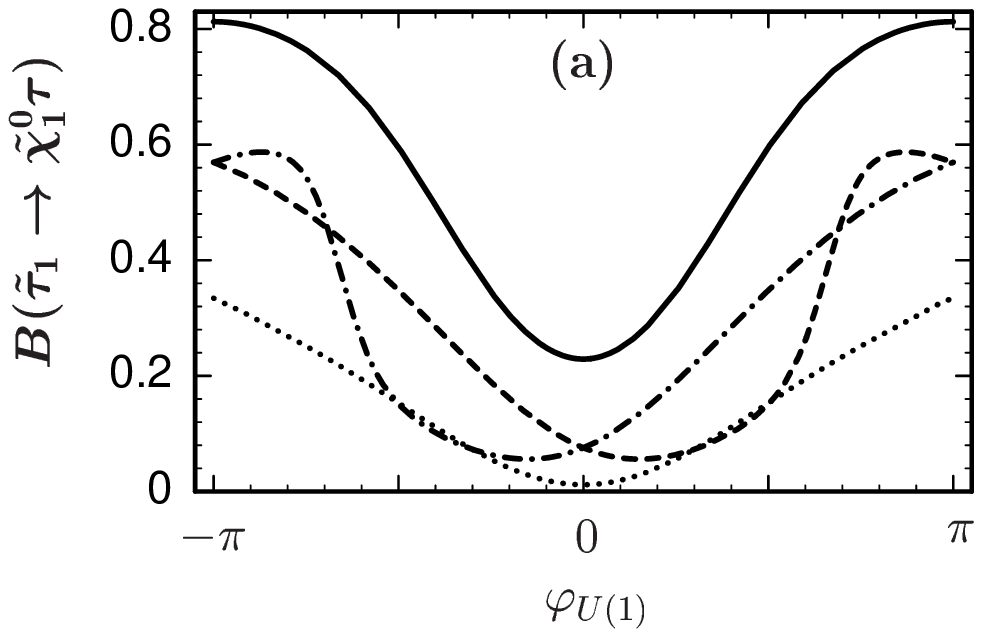,height=6.cm,width=8.cm}}}
\put(75,0){\mbox{\epsfig{figure=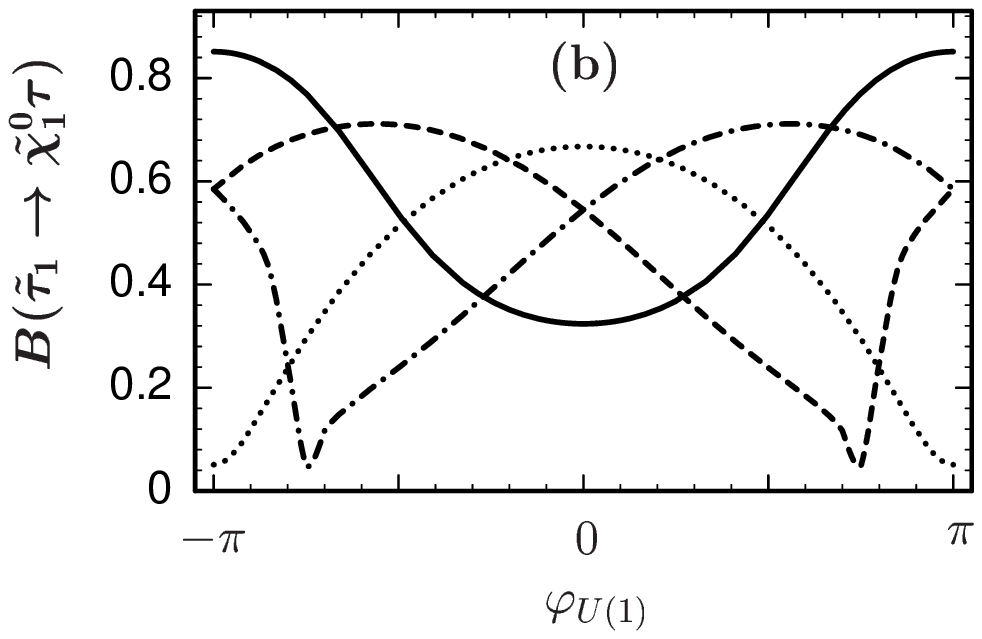,height=6.cm,width=8.cm}}}
\end{picture}
\end{center}
\caption{Branching ratio of $\T_1 \to \CH_1^0 \tau$ as a
function of $\varphi_{U(1)}$ for 
$\varphi_{\mu}=\pi (\rm solid~line)$, 
$\pi/2 (\rm dashed~line)$, $0 (\rm dotted~line)$, 
$-\pi/2 (\rm dashdotted~line)$, 
$\varphi_{A_{\tau}}=0$, 
$m_{\T_1}=240$~GeV, $m_{\T_2}=500$~GeV,
$|\mu|= 150$~GeV, $\tan\beta = 3$, and $|A_{\tau}| = 1000$~GeV, 
assuming a) $M_{\ti L} < M_{\ti E}$, $M_2=280$~GeV, 
b) $M_{\ti L} \geq M_{\ti E}$, $M_2=380$~GeV.}
\label{fig:stau1.4}
\end{figure}

In Figs.~\ref{fig:stau1.4}\,a,\,b  we show the $\varphi_{U(1)}$ 
dependence of $B(\T_1 \ra \CH^0_1\tau)$ for $|\mu|=150$~GeV, 
$\tan\beta=3$ and $\varphi_{A_{\tau}}=0$, for 
$\varphi_{\mu}=\pi$ (solid line), $\varphi_{\mu}=\pi/2$ (dashed
line), $\varphi_{\mu}=0$ (dotted line) and
$\varphi_{\mu}=-\pi/2$ (dash-dotted line).
In Fig.~\ref{fig:stau1.4}\,a we take $M_{\ti L} < M_{\ti E}$ and
$M_2=280$~GeV.
Fig.~\ref{fig:stau1.4}\,b is
for $M_{\ti L} \geq M_{\ti E}$ and $M_2=380$~GeV. 
Although the $\varphi_{U(1)}$ dependence of 
$B(\T_1 \ra \CH^0_1\tau)$ stems only from the 
$\varphi_{U(1)}$ dependence of the $\CH^0_{1,2}$ parameters, it is
quite pronounced. It is essentially explained by the 
$\varphi_{U(1)}$ dependences of $N_{11}$ and $N_{12}$, which 
enter in the couplings $f^{\tau}_{L1}$ and $f^{\tau}_{R1}$ (see 
Eqs.~(\ref{eq:coupl1}) -- (\ref{eq:coupl3})).
For example, the minimum of $B(\T_1 \ra \CH^0_1\tau)$ in 
Fig.~\ref{fig:stau1.4}\,b at $\varphi_{U(1)} \approx 3\pi/4 
~(-3\pi/4)$ for 
$\varphi_{\mu}=\pi/2 ~(-\pi/2)$ is caused by a corresponding
minimum of $|N_{11}|$.

\begin{figure}[H]
\begin{center}
\includegraphics{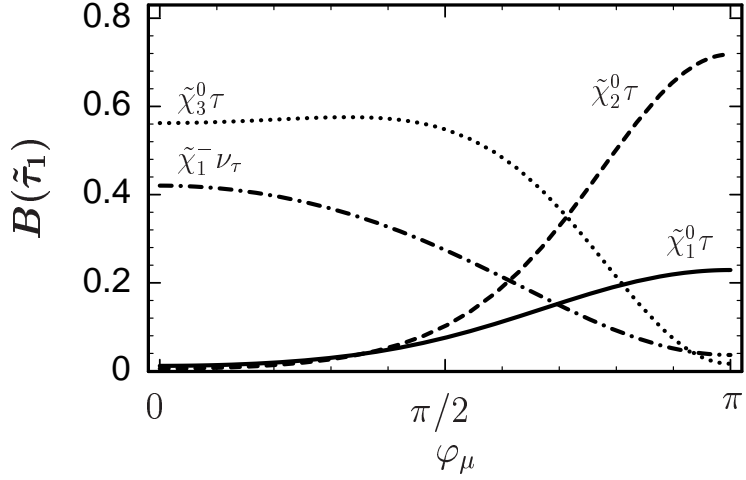}
\end{center}
\caption{Branching ratios of $\T_1 \to \CH^0_{1,2,3} \tau$ 
and $\T_1 \to \CH^-_1 \nu_{\tau}$ as a
function of $\varphi_{\mu}$ for 
$\varphi_{U(1)} = \varphi_{A_{\tau}} = 0$, 
$m_{\T_1}=240$~GeV, $m_{\T_2}=500$~GeV, 
$M_2=280$~GeV, $|\mu|=150$~GeV, $\tan\beta=3$, and $|A_{\tau}| =
1000$~GeV, assuming $M_{\ti L} < M_{\ti E}$.}
\label{fig:stau1.5}
\end{figure}

We have also studied how the branching ratios $B(\T_1 \ra
\CH^0_{2,3}\tau)$ and $B(\T_1 \ra \CH^-_1\nu_{\tau})$ vary as 
functions of the 
phases. As an example we show in Fig.~\ref{fig:stau1.5} 
these branching ratios as functions of $\varphi_{\mu}$ for
$\varphi_{U(1)} = \varphi_{A_{\tau}} = 0$, $M_2=280$~GeV,
$|\mu|=150$~GeV and  $\tan\beta=3$, assuming $M_{\tilde L} < M_{\tilde E}$. 
For this set of parameters all branching ratios
shown have a significant $\varphi_{\mu}$ dependence. Their
behaviour can be understood in the following way: If we first
consider $B(\T_1 \ra \CH^-_1 \nu_{\tau}) \propto |U_{11}|^2$,
the $\varphi_{\mu}$ dependence of $|U_{11}|$ follows from
\begin{equation}
|U_{11}|^2 = \cos^2\theta_1 = \frac{1}{2}\left(1+ \frac{|\mu|^2 - M_2^2 
+2 m_W^2 \cos2\beta}{m_{\CH^+_2}^2-m_{\CH^+_1}^2}\right),
\label{eq:U11}
\end{equation}
where $\theta_1$ is the mixing angle of the chargino mixing 
matrix $U_{ij}$ defined in Eq.~(\ref{eq:theta1}).
The mass squared difference $m_{\CH^+_2}^2-m_{\CH^+_1}^2$
decreases for $\varphi_{\mu} \to \pi$, which can be seen from
Eq.~(\ref{eq:massch}), therefore, also $|U_{11}|$ decreases. The
behaviour of $B(\T_1 \ra \CH^0_{1,2,3}\tau)$ can be understood
by noting that $\CH^0_{1,2,3}$ have large higgsino--components.
Varying $\varphi_{\mu}$ from $0$ to $\pi$ essentially interchanges the 
$\ti H^0_1$ and $\ti H^0_2$ components of $\CH^0_{1,2,3}$. This
causes the variation in the branching ratios, because $\T_1$ couples 
 to the $\ti H_1^0$ component
of $\CH^0_i$ but not to the $\ti H^0_2$ component.

It is expected that $\varphi_{\mu}$ and $\varphi_{U(1)}$
will be determined by measuring suitable observables of the 
chargino and neutralino sectors \cite{kalinow}. The 
$\varphi_{\mu}$ and $\varphi_{U(1)}$ dependences of the 
various $\T_1$ decay branching ratios, however, will give 
useful additional information for the precise determination of
$\varphi_{\mu}$ and $\varphi_{U(1)}$ and thereby provide further
tests of the MSSM with complex parameters. This may also be
helpful for resolving the ambiguities encountered in the 
studies about the parameter determination of the 
chargino and neutralino sectors \cite{kalinow}. 

An additional observable which is very sensitive to the SUSY
parameters of the $\T_i$ and ${\tilde \chi}^0_k$ systems is the 
longitudinal polarization of
the outgoing $\tau$--lepton in the decays
$\T_i \ra \CH^0_j\tau$ \cite{nojiri}. For the $\T_1$ decays into
neutralinos it is defined as
\begin{equation}
\mathcal{P}_{\tau}=\frac{B(\CH_j^0\tau_R)-B(\CH_j^0\tau_L)}
{B(\CH_j^0\tau_R)+B(\CH_j^0\tau_L)}
= \frac{|b_{1j}^{\T}|^2-|a_{1j}^{\T}|^2}
{|b_{1j}^{\T}|^2+|a_{1j}^{\T}|^2}
\label{eq:taupol}
\end{equation}
where the last equation holds in the limit
$m_{\tau} \ra 0$. $R, L$ denote $\lambda_{\tau}=+\frac{1}{2},
-\frac{1}{2}$, respectively.

\begin{figure}[H]
\setlength{\unitlength}{1mm}
\begin{center}
\begin{picture}(150,60)
\put(-5,0){\mbox{\epsfig{figure=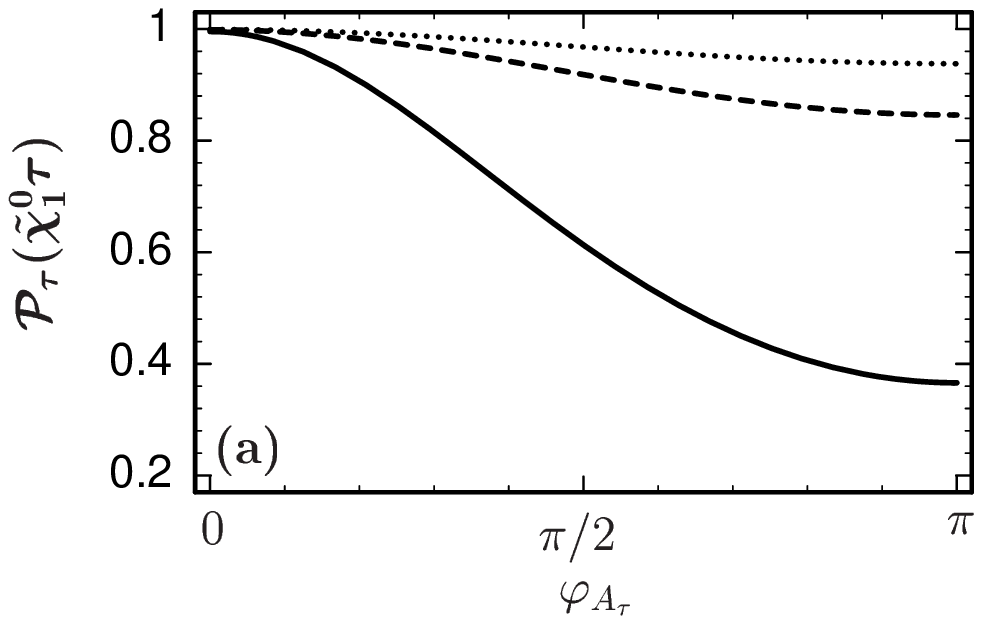,height=6.cm,width=8.cm}}}
\put(75,0){\mbox{\epsfig{figure=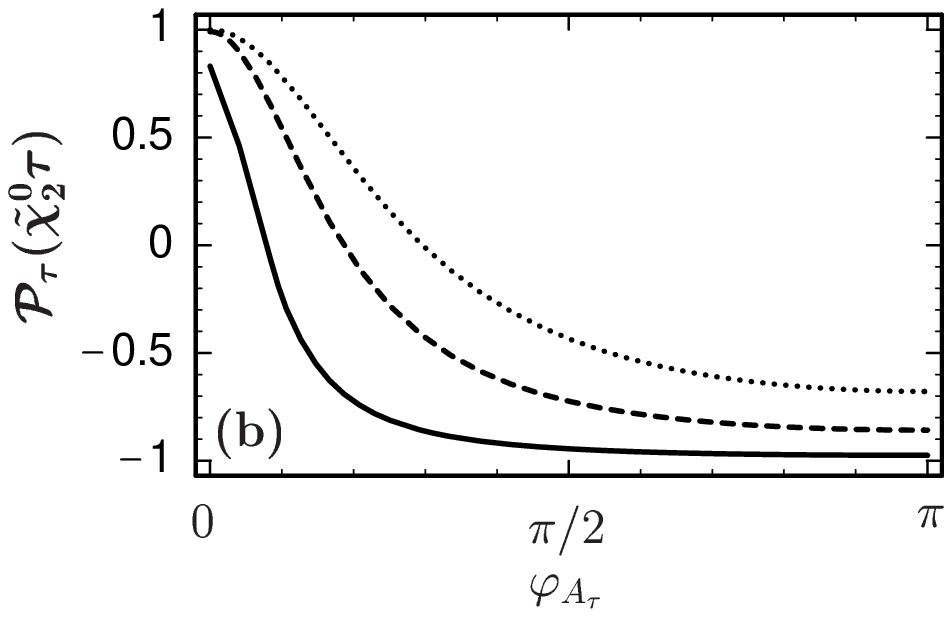,height=6.cm,width=8.cm}}}
\end{picture}
\end{center}
\caption{Longitudinal $\tau$ polarization, defined in
Eq.~(\ref{eq:taupol}), for a) $\T_1 \to \CH^0_1 \tau$ 
and b) $\T_1 \to \CH^0_2 \tau$ as a function of
$\varphi_{A_{\tau}}$. The parameters are 
$m_{\T_1}=240$~GeV, $m_{\SN}=233$~GeV $(\rm solid~line)$,
$238$~GeV $(\rm dashed~line)$, $243$~GeV $(\rm dotted~line)$, 
$\varphi_{\mu}= \varphi_{U(1)}=0$, $M_2=200$~GeV, 
$|\mu|=300$~GeV, $\tan\beta = 3$, and $|A_{\tau}|=1000$~GeV.} 
\label{fig:stau1.6}
\end{figure}

We show in Figs.~\ref{fig:stau1.6}\,a,\,b the
 longitudinal polarization of the $\tau$ in the decays
$\T_1 \ra \CH^0_1\tau$ 
and $\T_1 \ra \CH^0_2\tau$, respectively, as a function of 
$\varphi_{A_{\tau}}$ for $m_{\SN}=233$~GeV~(solid line), 
$238$~GeV~(dashed line) and $243$~GeV~(dotted line), which
correspond to $M_{\ti L}=240$~GeV, $245$~GeV and $250$~GeV,
respectively. 
The other parameters are $M_2=200$~GeV, 
$|\mu|=300$~GeV, $\tan\beta = 3$,
$\varphi_{\mu}= \varphi_{U(1)}=0$. The behaviour of 
$\mathcal{P}_{\tau}(\ti \chi^0_1 \tau)$ in
Fig.~\ref{fig:stau1.6}\,a follows from the change of the
mixing angle $\theta_{\ti \tau}$ with varying
$\varphi_{A_{\tau}}$, as described in the discussion of 
Fig.~\ref{fig:stau1.1}. The behaviour of 
$\mathcal{P}_{\tau}(\ti \chi^0_2 \tau)$ in
Fig.~\ref{fig:stau1.6}\,b can be understood by noting that 
in this case $\ti \chi^0_2$ is mainly a $\ti W^3$ which couples only to the
$\T_L$ component of $\T_1$ and that this component strongly increases
for $\varphi_{A_{\tau}} \to \pi$ as can be seen from 
Eqs.~(\ref{eq:rtau}) and (\ref{eq:thtau}).

\begin{figure}[H]
\begin{center}
\includegraphics{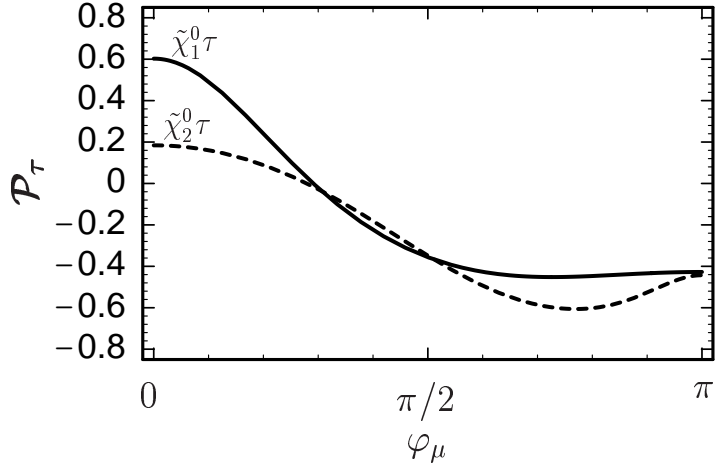}
\end{center}
\caption{Longitudinal $\tau$ polarization, defined in
Eq.~(\ref{eq:taupol}), for $\T_1 \to \CH^0_1 \tau$ 
and $\T_1 \to \CH^0_2 \tau$ as a function of
$\varphi_{\mu}$. The parameters are 
$\varphi_{U(1)} = \varphi_{A_{\tau}}=0$, 
$m_{\T_1}=240$~GeV, $m_{\T_2}=500$~GeV,
$M_2=350$~GeV, $|\mu|=150$~GeV,
$\tan\beta=3$, and $|A_{\tau}|=1000$~GeV,
assuming $M_{\ti L} < M_{\ti E}$.}
\label{fig:stau1.7}
\end{figure}

In Fig. \ref{fig:stau1.7} we show the longitudinal $\tau$ polarization
in the decays $\T_1 \to \CH^0_1\tau$ and $\T_1 \to \CH^0_2\tau$ as a
function of $\varphi_{\mu}$. Here we have taken $m_{\T_2}=500$~GeV and
the other parameters $M_2=350$~GeV, $|\mu|=150$~GeV, $\tan\beta=3$,
$\varphi_{U(1)} = \varphi_{A_{\tau}}=0$.  As we have chosen $M_{\ti L}
< M_{\ti E}$, $\T_1$ is mainly a $\T_L$ and $\mathcal{P}_{\tau}$ is
negative for $\varphi_{\mu} \gsim 3\pi/10$ due to the very small
$\tau$ Yukawa coupling.  For $\varphi_{\mu} \ra 0$, the $\T_L \tau_L
\tilde \chi^0_{1,2}$ couplings $|f^{\tau}_{L1}|$ and $|f^{\tau}_{L2}|$
decrease monotonically, because $\tilde \chi^0_{1,2}$
are mainly
higgsino-like and changing the phase $\varphi_{\mu}$ from $\pi$ to $0$ implies
essentially a decrease of their gaugino components as well as exchanging
the ${\tilde H}^0_1$ component with the ${\tilde H}^0_2$ component.  
This leads to a change of the sign of
$\mathcal{P}_{\tau}$. $|f^{\tau}_{L2}|$ has a maximum at
$\varphi_{\mu} \approx 3\pi/4$, which is clearly seen in the minimum
of $\mathcal{P}_{\tau}(\CH_2^0 \tau) \approx -0.6$ for this value of
$\varphi_{\mu}$. 

\subsection{$\T_2$ Decays}

As we have seen in the previous subsection, the branching
ratios for the fermionic $\T_1$ decays depend on the phase
$\varphi_{A_{\tau}}$  only via the
$\cos(\varphi_{A_{\tau}} + \varphi_{\mu})$ dependence of the
mass $m_{\T_1}$ and the mixing angle $\theta_{\T}$. We consider
now the bosonic $\T_2$ decays where the couplings to the Higgs
bosons explicitely depend on the phases $\varphi_{A_{\tau}}$ 
and $\varphi_{\mu}$ (see Eqs.~(\ref{chHcoup}) to
(\ref{eq:Hijcoup})). 
The decay widths into $W^{\pm}$, $Z$ and Higgs bosons are
enhanced by choosing $|\mu|$ and/or $|A_{\tau}|$ large \cite{bartl}.

As already mentioned, the RG improved 
Higgs sector is determined by the parameters 
$m_{H^{\pm}}$, $\tan\beta$, $|\mu|$, $|A_t|$, $|A_b|$, 
$\varphi_{\mu}$, $\varphi_{A_t}$, $\varphi_{A_b}$, 
$M_{\ti Q}$, $M_{\ti U}$, $M_{\ti D}$, $|m_{\ti g}|$, $\varphi_{\ti g}$,
$|M_1|$, $\varphi_{U(1)}$  and $M_2$ \cite{carenaetal}. 
We fix
$M_{\ti Q} = M_{\ti U} = M_{\ti D} = M_{SUSY}$. 
The amount of the $CP$ violating scalar--pseudoscalar transition
in the neutral Higgs mass matrix is proportional to the parameter
\begin{eqnarray}
\eta_{CP}=\frac{g^2 m^4_f |A_f| |\mu|}{128 \pi^2 m_W^2 M^2_{SUSY}}
\sin(\varphi_{\mu}+\varphi_{A_f}),
\end{eqnarray}
where $f=t,b$ \cite{carenaetal,pilaftsi1}. 
This means that significant $CP$ violating
effects in the Higgs sector can be expected if $|\mu|, |A_f| >
M_{SUSY}$ and $|\sin(\varphi_{\mu} + \varphi_{A_f})| \approx 1$.
As we focus on the $\varphi_{A_{\tau}}$ and the 
$\varphi_{\mu}$ dependence of the observables, we fix the
phases $\varphi_{A_t}=\varphi_{\ti g}=0$,
$\varphi_{A_b}=\pi$ and we  take 
$|A_t|=|A_b|=800$~GeV, $M_{SUSY}=600$~GeV, 
$|m_{\ti g}|=(\alpha_s(|m_{\ti g}|)/\alpha_2)M_2$ 
(with $\alpha_s(Q)=12\pi/((33-2n_f)\ln(Q^2/\Lambda^2_{n_f}))$,
$n_f$ being the number of quark flavors).
For this choice of parameters mixing between the CP--even and
CP--odd Higgs bosons at one loop level occurs only if
$\varphi_{\mu}\neq \{0,\pm\pi\}$. Therefore, we can control the
influence of explicit CP violation in the Higgs sector with the
parameter $\varphi_{\mu}$. 
With this choice of parameters the constraint
from the $\rho$-parameter on the $\ti t$ and $\ti b$
masses and mixings, $\delta\rho(\ti t-\ti b)<0.0012$, is always fulfilled
\cite{drees}. 

For large $\tan\beta$ the allowed range of $|\mu|$ is restricted
by the two-loop contributions to the EDMs of electron 
and neutron \cite{darwin}. For example, for $\tan\beta=40$,
$\varphi_{\mu}=\pi/2$, $m_{H^{\pm}} \lets 200$~GeV and the other
parameters as fixed above 
the EDMs give the restriction $|\mu| \lets 600$~GeV.
Therefore, we also fix $|\mu| = 600$~GeV. 

In the following we give some numerical examples which
show the dependence of the branching ratios for 
$\T_2\to\T_1 H_i, i= 1,2,3,$
on $\varphi_{A_{\tau}}$, $\tan\beta$ and $m_{H^{\pm}}$.
We take $|A_{\tau}| = 900$~GeV,
$M_2=450$~GeV and $\varphi_{U(1)}=0$. We  consider the 
case $M_{\ti L} > M_{\ti E}$, where $\T_2$ is mainly 
$\T_L$-like and $\T_1$ is mainly $\T_R$-like. In this case the
decays 
$\T_2 \to W^- \SN$ and $\T_2 \to H^- \SN$ are
kinematically forbidden.

\begin{figure}[H]
\begin{center}
\begin{picture}(150,60)
\put(-5,0){\mbox{\epsfig{figure=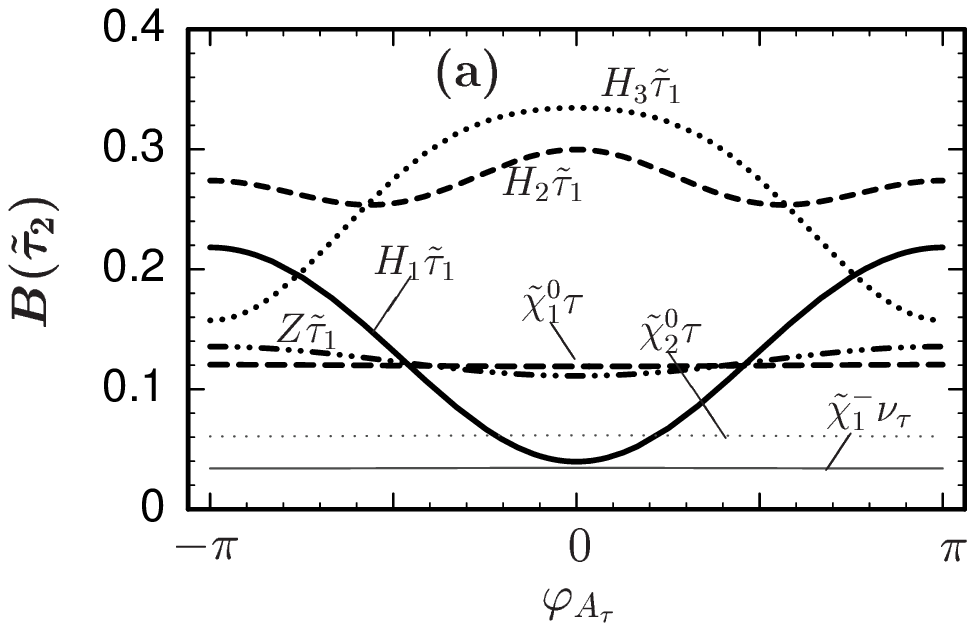,height=6.cm,width=8.cm}}}
\put(75,0){\mbox{\epsfig{figure=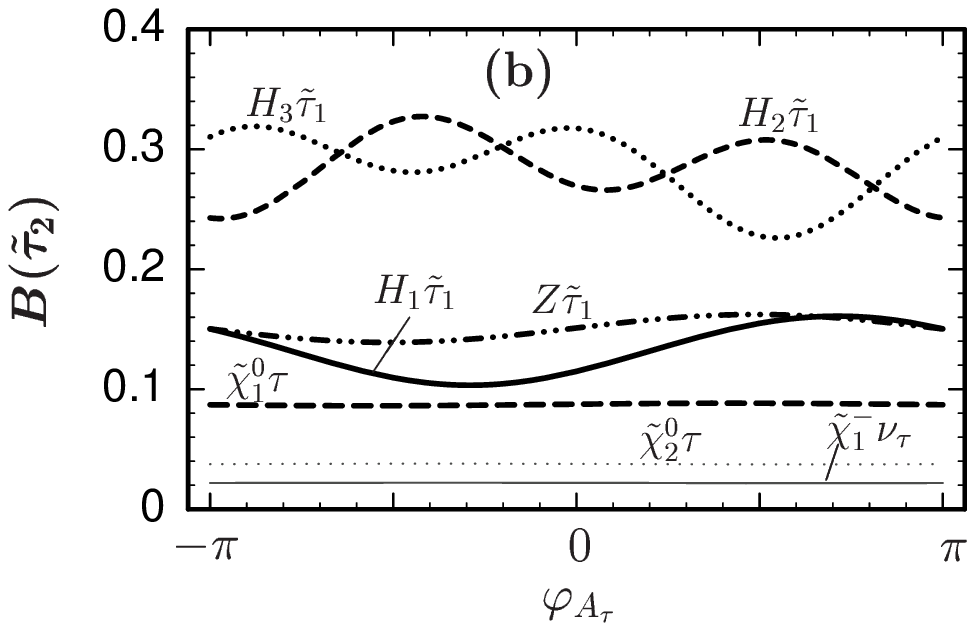,height=6.cm,width=8.cm}}}
\end{picture}
\end{center}
\caption{Branching ratios of $\T_2 \to H_{1,2,3} \T_1$, 
$\T_2 \to Z \T_1$, 
$\T_2 \to \CH^0_{1,2} \tau$ 
and $\T_2 \to \CH^-_1 \nu_{\tau}$ as a function of
$\varphi_{A_{\tau}}$ for a) $\varphi_{\mu}=0$ and b)
$\varphi_{\mu}=\pi/2$, with the other parameters 
$m_{\T_1}=240$~GeV, $m_{\T_2}=500$~GeV,
$m_{H^{\pm}}=160$~GeV, $|\mu|=600$~GeV, $M_2=450$~GeV, 
$\varphi_{U(1)}=0$, $\tan\beta=30$, and $|A_{\tau}| = 900$~GeV,
assuming $M_{\ti L} > M_{\ti E}$.}
\label{fig:stau1.9}
\end{figure}

In Figs.~\ref{fig:stau1.9}\,a,\,b we show the branching ratios 
for various fermionic and bosonic $\ti \tau_2$ decays as a
function of $\varphi_{A_{\tau}}$ for $\varphi_{\mu} = 0$ and
$\pi/2$, taking $\tan\beta = 30$,
$m_{H^{\pm}}=160$~GeV, $m_{\tilde \tau_1} =240$~GeV,
  $m_{\tilde \tau_2} =500$~GeV 
and the other parameters as specified
above. As can be seen, the branching ratios of the decays 
$\T_2 \to H_{1,2,3} \T_1$ show a pronounced 
($\varphi_{A_{\tau}}$,$\varphi_\mu)$ dependence. 
The behaviour of these branching ratios can be understood by
examining the approximate 
formula for the coupling squared for 
$\T_2\to\T_1 H_i,$
\begin{equation}\label{stau2coup1}
|C(\T_{2}^{\ast} H_i \T_{1})|^2\simeq|C(\T_{L}^{\ast} H_i \T_{R})|^2
\left(1-2 \sin^2\theta_{\T}
\cos^2\theta_{\T}( 1+\cos 2(\arg\lbrack C(\T_{L}^{\ast} H_i
\T_{R})\rbrack+\varphi_{\T}) 
)\right)
\end{equation}
with
\begin{eqnarray}\label{stau2coup2}
|C(\T_{L}^{\ast} H_i \T_{R})|^2 &\simeq & \frac{1}{2} Y_{\tau}^2
\biggl(\Bigl(|\mu|^2-|A_{\tau}|^2\Bigr)
\mathcal{O}_{2i}^2+|A_{\tau}|^2
-2|\mu||A_{\tau}|\mathcal{O}_{2i}\nonumber\\[3mm]
&&{}\times\Bigl(\mathcal{O}_{1i}
\cos(\varphi_{A_{\tau}}+\varphi_{\mu})-\mathcal{O}_{3i}
\sin(\varphi_{A_{\tau}}+\varphi_{\mu})\Bigr)\biggr),
\end{eqnarray}
which follows from Eq.~(\ref{eq:HLRcoup}) and (\ref{eq:Hijcoup}). Here we have 
omitted terms proportional to 
$(C(\T_{L}^{\ast} H_i \T_{L})-C(\T_{R}^{\ast} H_i \T_{R}))$
and $\cos\beta$. Eqs.~(\ref{stau2coup1}) and (\ref{stau2coup2}) 
show that a significant phase dependence of the 
$\T_2\to\T_1 H_i$ branching ratios can be expected for large
$\tan\beta$.
Moreover, also the $\varphi_\mu$ dependence of the Higgs mixing matrix elements
$\mathcal{O}_{ij}$ influences in a significant way 
the behaviour of $B(\T_2\to\T_1 H_i)$.
For $\varphi_{\mu} = 0$, for example, we obtain 
$\mathcal{O}_{11} \approx -0.262$, $\mathcal{O}_{21} \approx
-0.965$, $\mathcal{O}_{31} = 
\mathcal{O}_{12} = \mathcal{O}_{22} = 0$,
$\mathcal{O}_{32} = 1$, 
$\mathcal{O}_{13} \approx 0.965$, $\mathcal{O}_{23} \approx
-0.262$, $\mathcal{O}_{33} = 0$, 
$m_{H_1} = 115.74$~GeV, $m_{H_2} = 138.48$~GeV, 
$m_{H_3} = 139.14$~GeV. 
The $\varphi_{A_{\tau}}$ dependence of 
$B(\T_2 \to H_1 \T_1)$ follows essentially from the 
$\cos(\varphi_{A_{\tau}}+\varphi_{\mu})$ term and the first two
terms of Eq.~(\ref{stau2coup2}). The minimum of 
$B(\T_2 \to H_1 \T_1)$ at $\varphi_{A_{\tau}}=0$
(Fig.~\ref{fig:stau1.9}\,a) follows
from a partial cancellation of the terms in
Eq.~(\ref{stau2coup2}) (or, equivalently, from a partial cancellation of
the last two terms of Eq.~(\ref{eq:HLRcoup}), see also
Fig.~\ref{fig:stau1.11} below). 
The $\cos(\varphi_{A_{\tau}}+\varphi_{\mu})$ term and the first
two terms of Eq.~(\ref{stau2coup2}) determine also the
$\varphi_{A_{\tau}}$ behaviour of $B(\T_2 \to H_3 \T_1)$.
The $\varphi_{A_{\tau}}$ dependence of 
$B(\T_2 \to H_2 \T_1)$
follows from the last factor of Eq.~(\ref{stau2coup1}) and the
first term of Eq.~(\ref{eq:HLRcoup}).
As for Fig.~\ref{fig:stau1.9}\,b, 
for $\varphi_{\mu} = \pi/2$ we obtain 
$\mathcal{O}_{11} \approx -0.106$, $\mathcal{O}_{21} \approx
-0.992$, $\mathcal{O}_{31} \approx 0.066$,
$\mathcal{O}_{12} \approx -0.230$, $\mathcal{O}_{22} \approx
-0.040$, $\mathcal{O}_{32} \approx -0.972$, 
$\mathcal{O}_{13} \approx 0.967$, $\mathcal{O}_{23} \approx
-0.118$, $\mathcal{O}_{33} \approx -0.224$, 
$m_{H_1} = 117.09$~GeV, $m_{H_2} = 138.48$~GeV, $m_{H_3} =
139.14$~GeV. The $\varphi_{A_{\tau}}$ dependence of 
$B(\T_2 \to H_i \T_1)$ is
now different from that in Fig.~\ref{fig:stau1.9}\,a. In the
case of $B(\T_2 \to H_1 \T_1)$ the
$\cos(\varphi_{A_{\tau}}+\varphi_{\mu})$ term becomes 
$-\sin\varphi_{A_{\tau}}$ and it is multiplied by a much 
smaller factor, which explains the relatively flat 
$\varphi_{A_{\tau}}$ dependence. The behaviour of 
$B(\T_2 \to H_2 \T_1)$ and $B(\T_2 \to H_3 \T_1)$ can be
explained in an analogous way. 
For comparison we also plotted the branching ratios
of $\T_2 \to Z \T_1$ and of some of the decays into charginos
and neutralinos. 
The $\varphi_{A_{\tau}}$ dependence of $\T_2 \to Z \T_1$ 
essentially drops out (see Eq.~(\ref{eq:zkop})) and that of the
fermionic decays disappears due to the large value of
$\tan\beta$ for which $\theta_{\tilde \tau}$ is insensitive to
$\varphi_{A_\tau}$. 

We also studied the $\tan\beta$ dependence and the $m_{H^{\pm}}$
dependence of the $\T_2$ decay branching ratios into neutral
Higgs particles.
For $\tan\beta \to 0$ these branching ratios vanish
($\propto \tan\beta$), whereas for $\tan\beta > 10$ 
they depend only weakly on $\tan\beta$. The $m_{H^{\pm}}$
dependence of the branching ratio $B(\T_2 \to H_1 \T_1)$ is
shown in Fig.~\ref{fig:stau1.11} for 
$\varphi_{A_{\tau}} = 0, \pi/2, \pi$.
At $m_{H^{\pm}} = 150$~GeV and $\varphi_{A_{\tau}} = 0$ this 
branching ratio practically vanishes. The reason is that the
coupling $C(\T_L^{\ast} H_1 \T_R)$ practically vanishes for this
set of parameters due to a cancellation of the last two terms in
Eq.~(\ref{eq:HLRcoup}). At this point also a level crossing of
$H_1$ and $H_2$ occurs. We see that this branching ratio is sensitive to
$\varphi_{A_\tau}$ for $m_{H^\pm} \lsim 250$~GeV.

\subsection{$\ti\nu_{\tau}$ Decays}
 
The decay widths for $\ti\nu_{\tau}$ decays into charginos 
and neutralinos are independent of $\varphi_{A_{\tau}}$. The
decay widths for $\SN \to \ti \chi^+_k \tau$ depend on 
$\varphi_{\mu}$, those for $\SN \to \ti \chi^0_k \nu_{\tau}$
depend also on $\varphi_{U(1)}$.
We first assume $M_{\tilde L}<M_{\tilde E}$, which leads to a 
sneutrino mass $m_{\ti\nu_{\tau}} \simeq 229$~GeV for
$m_{\T_1}=240$~GeV, $m_{\T_2}=500$~GeV and $\tan\beta=3$.
In this case the decays $\SN \to W^+ \T_1$ and 
$\SN \to H^+ \T_1$ are kinematically forbidden. 

We show in Figs.~\ref{fig:stau1.8}\,a and b the
branching ratios for the 
decays into $\CH^0_1\nu_{\tau}$, $\CH^0_2\nu_{\tau}$ and
$\CH^+_1\tau$ as functions of $\varphi_{\mu}$ and
$\varphi_{U(1)}$, respectively, for $M_2=500$~GeV,
$|\mu|=150$~GeV, $\tan\beta=3$, and $|A_\tau|=1000$~GeV. 
In Fig.~\ref{fig:stau1.8}\,a
we take $\varphi_{U(1)}=0$ and in Fig.~\ref{fig:stau1.8}\,b
we take $\varphi_{\mu}=0$. 
As can be seen, the
branching ratio for  $\SN \to \CH^0_1\nu_{\tau}$ decreases for
$\varphi_{\mu}\to \pi$, whereas those for
$\SN \to\CH^0_2\nu_{\tau}$ and $\SN \to\CH^+_1\tau$ increase.
The decay widths $\Gamma(\SN\to\CH_1^0 \nu_{\tau})$ and 
$\Gamma(\SN\to\CH_2^0 \nu_{\tau})$ decrease for 
$\varphi_{\mu}\to \pi$,
because the matrix elements $|N_{12}|$ and $|N_{22}|$ decrease
for $\varphi_{\mu} \to \pi$. 
The matrix element $|V_{11}|$ entering the decay width
$\Gamma(\SN\to\CH_1^+\tau)$ also decreases, see Eqs.~(\ref{eq:Vchar}) and 
(\ref{eq:theta2}). However, as
$\Gamma(\SN\to\CH_1^+\tau)$ and $\Gamma(\SN\to\CH_2^0 \nu_{\tau})$ decrease 
more slowly than the total decay width,
 the corresponding branching ratios increase
for $\varphi_{\mu}\to \pi$.
In Fig.~\ref{fig:stau1.8}\,b the branching ratio
$B(\ti\nu_{\tau} \to \CH_1^0 \nu_{\tau})$ 
decreases for $\varphi_{U(1)}\to \pi$ and 
$B(\SN \to \CH_1^+ \tau)$
increases. The reason is that $|N_{11}\tan \Theta_W -N_{12}|$ 
and hence the width $\Gamma(\SN \to\CH_1^0 \nu_{\tau})$ rapidly decreases for
$\varphi_{U(1)}\to \pi$. 
$B(\SN \to \CH_1^+ \tau)$ increases due to the decrease of the
total decay width. 

In the case $M_{\ti L} - M_{\ti E} \grts m_W,  m_{H^+}$ 
also the bosonic decays 
$\SN \ra  W^+ \T_1, H^+ \T_1$ are kinematically allowed.
Consequently, the branching ratios of the fermionic decays are
reduced.  It turns out that in most cases the bosonic decay widths are almost 
independent of the phases; only in the region $|\mu| \ll m_{\SN} < |M_{1,2}|$
a significant dependence on the phases is possible. 
For small $\tan \beta$ the phase dependence of the width 
$\Gamma(\SN \ra H^+ \T_1)$ tends to be suppressed, because of the small
Yukawa coupling, see Eq.~(\ref{chHcoup}). For large $\tan\beta$ the term
$m_{\tau} A_{\tau}^* \tan\beta$ in Eq.~(\ref{chHcoup}) dominates
and $\Gamma(\SN \ra  H^+ \T_1) \propto$ 
$|\sin \theta_{\tilde \tau} C(\SN^{\ast} H^+ \T_{R})|^2 \propto
 |\sin \theta_{\tilde \tau} m_\tau A^*_\tau \tan\beta|^2$ becomes essentially
independent of the phases. Note here that $\theta_{\tilde \tau}$ is 
hardly sensitive to the phases because 
$M^2_{\ti L} - M^2_{\ti E} \gg m_\tau |A_\tau - \mu^*\tan\beta|$ in these
scenarios.
The phase dependence of the width 
$\Gamma(\SN \ra W^+ \T_1)$ is caused only by the phase
dependence of $\cos\theta_{\ti \tau}$ (see Eq.~(\ref{eq:wkop})) and is
again weak by the same reasoning as above.

\section{Parameter Determination}
 
 We now study the extent to which one can extract
the underlying parameters from measured masses, branching ratios and
cross sections. 
In the following we assume that an integrated luminosity of 2 ab$^{-1}$ is
available. At a high luminosity collider like TESLA one can expect
that this amount of integrated luminosity will be accumulated in four
years of running \cite{tdr}.
Our strategy is as follows: 
\begin{enumerate}
\item Take a specific set of values of the MSSM
parameters. 
\item Calculate the masses of
$\tilde \tau_i$, $\tilde \chi^0_j$, $\tilde \chi^\pm_k$, the 
production cross sections for 
$e^+ e^- \to \tilde \tau_i \overline{\tilde \tau}_j$
and branching ratios of 
the $\T_i$ decays. 
\item Regard these calculated values as real experimental data with definite
errors.
\item Determine the underlying MSSM parameters and their errors from the
``experimental data'' by a fit. 
\end{enumerate}
We have checked that inclusion of the data on the mass, production and
decays of $\tilde \nu_\tau$ does not further improve the accuracy of
the underlying parameters to be determined. The reason is that the
expected relative errors of the data in the sneutrino sector are
larger than those in the stau sector \cite{Baer2,Martyn2}.

We have taken the following input
parameters for the calculation of these observables:
$M_{\ti E} =$ 150~GeV, $M_{\ti L} =$ 350~GeV, 
$A_\tau =$ -800 i GeV, $M_2 =$ 280~GeV, 
$\mu =$ 250~GeV and $\varphi_{U(1)}=0$. We have considered the cases
$\tan \beta = 3$ and $30$. The Higgs sector has been fixed with 
$m_{H^+} = 170$~GeV (160), $m_{A^0} = 151.4$~GeV (138.5), 
$m_{h^0} = 113.3$~GeV (115.7),
$m_{H^0} = 155.6$~GeV (139.1) and $\sin \alpha = 0.432$ (-0.26)
in case of $\tan \beta=3$ (30). Here $h^0$, $H^0$, $A^0$, $\alpha$ are
the lighter CP-even Higgs boson, the heavier CP-even Higgs boson, the
CP-odd Higgs boson and the mixing angle of the CP-even Higgs bosons,
respectively.
Here we focus on the
determination of the phase $\varphi_{A_{\tau}}$ of $A_{\tau}$,
therefore, we neglect mixing of the $CP$-even and $CP$-odd Higgs
states. 
We have taken the relative errors of stau masses,
chargino and neutralino masses from \cite{tdr,Martyn:1999tc}, which
we rescale according to our scenario;
in case of $\tan \beta = 30$ we have taken into account an
additional factor of 3 for the errors (relatively to $\tan\beta=3$) due to the
reduced efficiency in case of multi $\tau$ final states as
indicated by the studies in \cite{Nojiri:1996fp}.
We take the errors of the Higgs 
 mass parameters as $\Delta m_{h^0} = 50$~MeV, $\Delta
 m_{H^0} =\Delta m_{A^0} =\Delta m_{H^+} = 1.5$~GeV \cite{tdr} for
$\tan\beta=3$ and 30.  
For the branching ratios and the production cross sections we
have taken the statistical errors only. 
We give the values of the calculated masses 
and assumed errors in Table~\ref{tab:masserrors} and those of the
calculated branching ratios of $\tilde \tau_2$ decays in
Table~\ref{tab:branchingratios}. 
$\tilde \tau_1$ decays only into $\tau \ti\chi^0_1$ for both
values of $\tan \beta$, because this is the only channel open.
  
\begin{table}
 \caption{Calculated masses and their assumed errors (in GeV).}
 \label{tab:masserrors}
\begin{center}
 \begin{tabular}{|c|c||c|c|}
 \hline
 \multicolumn{2}{|c||}{$\tan \beta=3$} &\multicolumn{2}{c|}{$\tan \beta=30$} \\
 \hline
 $m_{\tilde \tau_1} = 155.0 \pm 0.7$ & $ m_{\tilde \tau_2}=352.6 \pm 1.2$ & 
 $m_{\tilde \tau_1} = 150.6 \pm 2.1$ & $ m_{\tilde \tau_2}=355.7 \pm 3.6$ \\ 
 $m_{\tilde \chi^0_1}=125.6 \pm 0.17$& $m_{\tilde \chi^0_2}=205.6\pm0.11$ &
 $m_{\tilde \chi^0_1}=133.2 \pm 0.56$& $m_{\tilde \chi^0_2}=214.3\pm0.35$ \\
 $m_{\tilde \chi^0_3}=253.5 \pm 0.24$& $m_{\tilde \chi^0_4}=343.1\pm0.51$ &
 $m_{\tilde \chi^0_3}=258.0 \pm 0.73$& $m_{\tilde \chi^0_4}=331.4\pm1.4$ \\
 $m_{\tilde \chi^+_1}=194.0 \pm 0.06$& $m_{\tilde \chi^+_2}=340.9\pm0.25$ &
 $m_{\tilde \chi^+_1}=210.0 \pm 0.19$& $m_{\tilde \chi^+_2}=331.6\pm0.72$ \\
\hline
 \end{tabular}
\end{center}
\end{table}
  
\begin{table}
 \caption{Branching ratios of $\tilde \tau_2$ decays calculated for
           $M_{\ti E} =$    150~GeV,
           $M_{\ti L} =$    350~GeV,
           $A_\tau =$  -800 i GeV,
           $M_2 =$  280~GeV,
           $\mu =$  250~GeV and $\varphi_{U(1)}=0$. 
           We show only branching ratios larger than $10^{-3}$.}
 \label{tab:branchingratios}
\begin{center}
 \begin{tabular}{|c||c|c|c|c|c|c||c|c|c|c|}
\hline
$ \tan \beta$ & $\tau \tilde \chi^0_1$ & $\tau \tilde \chi^0_2$
              & $\tau \tilde \chi^0_3$ & $\tau \tilde \chi^0_4$
              & $\nu_\tau \tilde \chi^-_1$ & $\nu_\tau \tilde \chi^-_2$
              & $ Z \tilde \tau_1$ & $ A^0 \tilde \tau_1$
              & $ h^0 \tilde \tau_1$ & $ H^0 \tilde \tau_1$ \\
\hline
3 & 0.116 & 0.423 & 0.001 & 0.002 & 0.438 & 0.008  
  & 0.002 & 0.008 & 0.003 & 0 \\
30 & 0.107 & 0.195 & 0.036 & 0.008 & 0.135 & 0.019
  & 0.044 & 0.393 & 0.062 & 0.001 \\
\hline
 \end{tabular}
\end{center}
 \end{table}

For the determination of the stau parameters we have used the
information obtained from the measurement of the stau masses at 
threshold and the production cross sections of $\T_i
\bar{\T_j}$ pairs at $\sqrt{s}=800$~GeV for two
different $(e^-,e^+)$ beam polarizations $(P_-,P_+) = (0.8, -0.6)$ and
$(P_-,P_+) = (-0.8, 0.6)$. Here we have assumed that 
a total effective luminosity of 250 fb$^{-1}$ is avaible for
each choice of polarization. The cross section measurements are important
for the determination of $|\cos \theta_{\tilde \tau}|^2$ as can
be seen from Eq. (\ref{eq:zkop}) and the formulae for the cross
sections in \cite{sferm}. 
 In addition we have used the information from all 
branching ratios in Table~\ref{tab:branchingratios} 
(with corresponding statistical errors).
These branching ratios together with the masses and cross
sections form an over--constraining system of observables for the
parameters 
$M_{\ti L}$, $M_{\ti E}$, $\Re e A_{\tau}$, 
$\Im m A_{\tau}$, $\Re e \mu$,
$\Im m \mu$, $\tan\beta$, $\Re e M_1$,
$\Im m M_1$, $M_2$.
We have determined these parameters and their errors from the ``experimental
data'' on these observables by a
least--square fit.
The results obtained are shown in Table~\ref{tab:error}. As one 
can see, all parameters can be determined rather precisely.
$\tan\beta$ can be determined with
an accuracy of about 2\% in the case of $\tan\beta=30$ and about 1\% in
the case of $\tan\beta=3$. The relative error of the remaining parameters
except $A_\tau$
is about 1\%.
For $A_{\tau}$ we obtain the errors
$\Delta \Im m A_{\tau}/|A_{\tau}| \approx 9\%$, 
$\Delta \Re e A_{\tau}/|A_{\tau}| \approx 22\%$ in the case
$\tan\beta = 3$, and
$\Delta \Im m A_{\tau}/|A_{\tau}| \approx 3\%$, 
$\Delta \Re e A_{\tau}/|A_{\tau}| \approx 7\%$ in the case
$\tan\beta = 30$.
At first glance it might
be surprising that the errors of the stau parameters are
relatively small in case of large $\tan\beta$, despite the fact
that the assumed 
errors of the masses are larger for large $\tan\beta$. The error of
$A_\tau$ even decreases. The reason for this is the large branching
ratio for $\tilde \tau_2 \to A^0 \tilde \tau_1$ in the case
$\tan\beta = 30$ 
and the input parameters chosen (see Table~\ref{tab:branchingratios}), which 
gives a strong constraint on $|A_\tau|$.
For the determination of $A_{\tau}$ it is important that the
$\tilde \tau_2$ decays into neutral Higgs bosons are 
kinematically allowed, because their couplings to the staus are
practically proportional to $A_\tau \tan \beta$. 
Otherwise one would have to include the
decays of the heavier Higgs bosons to get  additional information
on $A_\tau$ from their decays into staus. This will be discussed
in a forthcoming paper \cite{futurework}. Additional information
could also be obtained at a $\mu^- \mu^+$ collider. In case of
sizable CP violating phases 
${\tilde \tau}_1 \overline{{\tilde \tau}}_1$ pairs can be
produced at the resonances of both heavier neutral Higgs states $H_{2,3}$
\cite{drees2} whereas in case of CP conservation 
$\tilde \tau_1 \overline{{\tilde \tau}}_1$
 pairs can only be produced at the $H^0$ resonance but not at the
$A^0$ resonance \cite{Bartl:1998qc}.

In the procedure described above we have determined the errors of the
fundamental parameters assuming an  integrated luminosity of 2 ab$^{-1}$, 
taking the expected experimental errors of the masses
from the Monte Carlo studies in 
\cite{tdr,Martyn:1999tc} and rescaling them to our scenario.
It is clear that further detailed Monte Carlo studies including
experimental cuts and detector simulation are necessary to determine more
accurately the expected experimental errors of the observables
for our scenario, in
particular the errors of the stau decay branching ratios.
Such a study is, however,
beyond the scope of this paper. Instead we have studied how our results
for the errors of the fundamental parameters are changed when the
experimental errors of the various observables are changed: we have redone
the procedure doubling the errors of the masses and/or
branching ratios and/or cross sections. Clearly we have found that the errors
of all parameters are approximately doubled if all experimental
errors are doubled.
Moreover, in this way we can see to which observables an individual parameter
is most sensitive.
Concentrating on the stau sector we find that the precision
of $M^2_{\ti E}$ and  $M^2_{\ti L}$ is sensitive to the stau mass 
determination
at the threshold as well as to the measurement of the total cross sections
in the continuum. The accuracy of $A_\tau$ is most sensitive to precise
measurements of the branching ratios, especially to those for the 
decays into Higgs bosons. The precision of $\mu$ is more sensitive to the 
errors
of chargino and neutralino masses than to the errors of the stau observables.
In the case of large $\tan \beta$, the precision of $\tan \beta$ depends 
significantly
on the precision of the stau cross sections and to a lesser extent
also on that of the stau decay branching ratios. 

In our procedure we have also determined the expected errors of $\Re e \mu$,
$\Im m\mu$, $\tan\beta$, $\Re e M_1$, $\Im m M_1$, $M_2$ 
using also the information
obtainable from  mass measurements of charginos and neutralinos.
As one can see in Table~\ref{tab:error}, the results are
quite satisfactory. 
Once these parameters together with the Higgs mass and mixing
parameters are precisely determined in the chargino, neutralino
and Higgs sectors, one can then
include them as input values in the determination of the parameters of
the stau sector. This will in turn improve the accuracy in the
determination of $\Re e(A_\tau)$ and $\Im m(A_\tau)$. Note that this
accuracy of the paramters at the weak scale 
allows also a rather precise determination of parameters at
a high scale, e.g.~the GUT scale, and hence 
the reconstruction of the parameters
of an underlying theory at this high scale \cite{Blair:2000gy}.

 \begin{table}
 \caption{Extracted parameters from the ``experimental data'' of the
  masses, production cross sections and decay branching ratios of
   $\tilde \tau_i$. The
original parameter point is specified by:
           $M_{\ti E} =$    150~GeV,
           $M_{\ti L} =$    350~GeV,
           $A_\tau =$  -800 i GeV,
           $M_2 =$  280~GeV,
           $\mu =$  250~GeV and $\varphi_{U(1)}=0$.}
 \label{tab:error}
\begin{center}
 \begin{tabular}{|l|c|c|}
 \hline
 $\tan \beta$ & 3  & 30 \\
 \hline
   $M^2_{\ti E}$~[GeV$^2$] &   2.25 $\cdot 10^4 \pm$  2.2 $\cdot 10^2$
           &   2.25 $\cdot 10^4 \pm$  6.0 $\cdot 10^2$ \\
   $M^2_{\ti L}$~[GeV$^2$] &   1.225 $\cdot 10^5 \pm$   4.3 $\cdot 10^2$
           &   1.229 $\cdot 10^5 \pm$   7.0 $\cdot 10^2$ \\
   $\Re e(A_\tau)$~[GeV] &   -8.0 $\pm$  180  &   8.0 $\pm$  55  \\
   $\Im m(A_\tau)$~[GeV] &  -800  $\pm$  70 &  -800  $\pm$  21  \\
 \hline
   $\Re e(\mu)$~[GeV] &   249.9 $\pm$   0.26 &   249.9 $\pm$   0.6 \\
   $\Im m(\mu)$~[GeV] & 2.4 $\pm$  1.7 &  -0.2 $\pm$  3.8 \\
   $\tan \beta$ &   2.999 $\pm$  2.7 $\cdot 10^{-2}$ &   29.9 $\pm$  0.70 \\
 \hline
   $\Re e(M_1)$~[GeV] &   140.9  $\pm$  0.21 &   140.6  $\pm$  0.63 \\
   $\Im m(M_1)$~[GeV] &   -0.7  $\pm$  3.4  &   0.16  $\pm$  1.0 \\
   $M_2$~[GeV] &   280  $\pm$  0.29 &   280  $\pm$  1.0 \\
 \hline
 \end{tabular}
\end{center}
 \end{table}

\section{Electric Dipole Moment of the $\tau$-lepton}

The MSSM with complex parameters  implies also a possible electric dipole
moment (EDM) of the $\tau$-lepton, which is induced by
chargino--sneutrino as well as stau--neutralino loops. For the
calculation of the $\tau$ EDM 
we use the corresponding formulae given in \cite{EDM2} for
the electron EDM by replacing $m_e$ by $m_\tau$. It turns out
that the natural range for the $\tau$ EDM is $O(10^{-22})
-O(10^{-21})$ $e$cm. This is demonstrated in Fig.~\ref{fig:edm} where 
we show the $\tau$ EDM $d_\tau$ corresponding to
some of the scenarios discussed above.
This is about 5--6 orders of magnitude below the current
experimental limit: $|d_{\tau}^{exp}| < 3.1 \cdot 10^{-16}$ $e$cm \cite{PDG}.

The dominant contribution stems from
the chargino loops as in case of electrons. 
However, for the $\tau$ EDM the neutralino loop is much more
important than in case of the electron due to the fact that $m_\tau
\gg m_e$.  Its modulus can reach about 10\% of the chargino--loop contributions
as can be seen in Figs.~\ref{fig:edm}b and f. The solid line shows
the total $\tau$ EDM, the dashed line the chargino--loop contributions and
the dotted line the neutralino--loop contributions. In the other plots of
Fig.~\ref{fig:edm}
the $\tau$ EDM is identical to the neutralino--loop
contributions, because in these scenarios $\varphi_\mu=0$ and hence the
chargino--loop contribution vanishes.

\section{Summary}
In this paper we have presented a phenomenological study of
$\tau$--sleptons $\T_i$ and $\tau$--sneutrinos $\SN$ 
in the Minimal Supersymmetric Standard Model with complex
parameters $A_{\tau}$, $\mu$ and $M_1$. We have taken into
account explicit $CP$ violation in the Higgs sector induced by
$\ti t_i$ and $\ti b_i$ loops with complex $\mu$ and complex
trilinear coupling parameters $A_t$ and $A_b$.
We have analysed production and decays of the $\T_i$ and $\SN$
 at a future   $e^+e^-$ linear collider. 
We have presented numerical predictions for the
fermionic and bosonic decays of $\T_{1}$, $\T_{2}$ and $\SN$.
We have analyzed their
SUSY parameter dependence, paying particular attention to their
dependence on the phases $\varphi_{A_{\tau}}$, $\varphi_{\mu}$
and $\varphi_{U(1)}$. For $\tan\beta \lets 10$ the phase dependence of
the branching ratios of the fermionic decays of $\T_1$ and $\SN$ is
significant whereas it becomes less pronounced for
$\tan\beta > 10$. The branching ratios of the $\T_2$ decays into
Higgs bosons depend very sensitively on the phases if 
$\tan\beta \grts 10$. Quite generally one can say that 
the decay pattern of the $\T_i$ and $\SN$
becomes even more involved if the parameters $A_{\tau}$, $\mu$
and $M_1$ are complex and if mixing of the $CP$-even and
$CP$-odd Higgs bosons is taken into account.

We have also given an estimate of the expected accuracy
 in the determination of the MSSM
parameters of the $\T_i$ sector by measurements of the
 masses, branching ratios and cross sections. We have considered
the cases $\tan\beta = 3$ and $\tan\beta = 30$. 
We have found that on favorable conditions the accuracy of the parameter 
$A_\tau$ can be expected to
be of the order of 10\% and that of the remaining stau parameters 
in the range of approximately 1\% to 3\%,
assuming an integrated luminosity of 2 ab$^{-1}$.
In addition we have considered  the electric dipole moment of
the $\tau$--lepton induced by the complex parameters in the
stau sector as well as the chargino and neutralino sectors. 
We find that it is well below the current experimental limit.

\section*{Acknowledgements:}
We thank A.~Pilaftsis and C.E.M.~Wagner for clarifying discussions and
correspondence. Furthermore, we are very grateful to H. Eberl
and S.~Kraml for valuable 
discussions and help in the numerical calculations, and to 
M.~Drees, W.~Majerotto and H.-U.~Martyn for useful discussions. 
This work was supported by the 
`Fonds zur F\"orderung der wissenschaftlichen Forschung'
of Austria FWF, Project No. P13139-PHY, by the Spanish DGICYT grant PB98-0693,
by Acciones Integradas Hispano--Austriaca
and by the European
Community's Human Potential Programme under contracts HPRN-CT-200-00148 and
HPRN-CT-2000-00149. 
T.K. is supported by a fellowship of the European Commission
Research Training Site contract HPMT-2000-00124 of the host group.
W.~P.~is supported by the 'Erwin
Schr\"odinger fellowship No.~J2095' of the `Fonds zur
F\"orderung der wissenschaftlichen Forschung' of Austria FWF and
partly by the Swiss `Nationalfonds'.

\begin{appendix}

\section{Chargino Masses and Mixing}
\label{app:char}

The chargino mass matrix in the weak basis is given by \cite{nilles,guha}
\begin{equation}
\mathcal{M}_C=\left(\begin{array}{ccc}
M_2 & \sqrt{2}m_W s_{\beta}\\[4mm]
\sqrt{2}m_W c_{\beta} & |\mu| e^{i\varphi_{\mu}}
\end{array}\right).
\end{equation}
$M_2$ is the $SU(2)$ gaugino mass parameter.
$c_{\beta}$ and $s_{\beta}$ are shorthand notations for 
$\cos \beta$ and $\sin \beta$, respectively.
This complex $2 \times 2$ matrix is diagonalized by the unitary 
$2 \times 2$ matrices $U$ and $V$:
\begin{equation}\label{dia}
U^{\ast} \mathcal{M}_C V^{\dag}=
diag(m_{\tilde\chi_1^{\pm}},m_{\tilde\chi_2^{\pm}})
, \hspace{2cm} 0\le m_{\tilde\chi_1^{\pm}} \le m_{\tilde\chi_2^{\pm}}.
\label{eq:massCH}
\end{equation}
The unitary matrices $U$ and $V$ can be parameterized in the
following way: 
\begin{equation}
U=
\left(\begin{array}{ccc}
e^{i\gamma_1} & 0\\[4mm]
0 & e^{i\gamma_2}
\end{array}\right)
\left(\begin{array}{ccc}
\cos\theta_1 & e^{i\phi_1}\sin\theta_1\\[4mm]
- e^{-i\phi_1}\sin\theta_1 & \cos\theta_1
\label{eq:Uchar}
\end{array}\right)
\end{equation}
\begin{equation}
V=
\left(\begin{array}{ccc}
\cos\theta_2 & e^{-i\phi_2}\sin\theta_2\\[4mm]
- e^{i\phi_2}\sin\theta_2 & \cos\theta_2
\end{array}\right)
\label{eq:Vchar}
\end{equation}
with
\begin{equation}
\tan2\theta_1=\frac{2\sqrt{2}m_W\lbrack M_2^2 c^2_{\beta}+
|\mu|^2 s^2_{\beta}+M_2|\mu| \sin2\beta\CPM \rbrack^{1/2}}{M_2^2- 
|\mu|^2-2 m^2_W \cos2\beta}
\label{eq:theta1}
\end{equation}
\begin{equation}
\tan2\theta_2=\frac{2\sqrt{2}m_W\lbrack M_2^2 s^2_{\beta}+
|\mu|^2 c^2_{\beta}+M_2|\mu| \sin2\beta\CPM
\rbrack^{1/2}}{M_2^2- 
|\mu|^2+2 m^2_W \cos2\beta}
\label{eq:theta2}
\end{equation}
\begin{equation}
\tan\phi_1=\sin{\varphi_{\mu}}\left(\CPM+
\frac{M_2 \cot\beta}{|\mu|}\right)^{-1} 
\label{eq:phi1}
\end{equation}
\begin{equation}
\tan\phi_2=-\sin{\varphi_{\mu}}\left(\CPM+
\frac{M_2 \tan\beta}{|\mu|}\right)^{-1}
\label{eq:phi2}
\end{equation}
\begin{equation}
\tan\gamma_1=-\sin{\varphi_{\mu}}\left(\CPM+
\frac{M_2 (m^2_{\CH^{\pm}_1}-|\mu|^2)}
{|\mu|m^2_W \sin2\beta}\right)^{-1} 
\label{eq:gam1}
\end{equation}
\begin{equation}
\tan\gamma_2=\sin{\varphi_{\mu}}\left(\CPM+
\frac{M_2 m^2_W \sin2\beta}
{|\mu|(m^2_{\CH^{\pm}_2}-M_2^2)}\right)^{-1} 
\label{eq:gam2}
\end{equation}
where $- \pi/2 \leq\theta_{1,2}\leq 0$.
The mass eigenvalues squared are
\begin{eqnarray}
m_{\tilde{\chi}_{1,2}^+}^2 &=& \frac{1}{2}
\biggl(M_2^2+|\mu|^2+2m_W^2\mp\Bigl((M_2^2-|\mu|^2)^2+4m_W^4 \cos^2 2\beta
+4 m_W^2(M_2^2+|\mu|^2 {}\nonumber\\[4mm]
&&{}+2 M_2|\mu| \sin 2\beta
\cos\varphi_{\mu})\Bigr)^{\frac{1}{2}}\biggr).
\label{eq:massch}
\end{eqnarray}

\section{Neutralino Masses and Mixing}
\label{app:neut}

The neutralino mass matrix in the weak basis $(\tilde B, 
\tilde W^3, \tilde H_1^0, \tilde H_2^0)$ is given as \cite{nilles,guha}:
\begin{equation}
\mathcal{M}_N = \ \left( \begin{array}{cccc}
|M_1| e^{i\varphi_{U(1)}} & 0 & -m_Z s_W c_{\beta} & m_Z
s_W s_{\beta}\\[3mm]
0 & M_2 & m_Z c_W c_{\beta} & -m_Z c_W s_{\beta}\\[3mm]
-m_Z s_W c_{\beta} & m_Z c_W c_{\beta} & 0 &
-|\mu| e^{i\varphi_{\mu}}\\[3mm]
m_Z s_W s_{\beta} & -m_Z c_W s_{\beta} &
-|\mu| e^{i\varphi_{\mu}} & 0
\end{array}\right),
\label{eq:massN}
\end{equation}
\\    
where $M_1$ is $U(1)$ gaugino mass parameter, 
with $\varphi_{U(1)}$ being the phase of $M_1$;
$c_W$ and $s_W$ are shorthand notations
for $\cos \Theta_W$ and $\sin \Theta_W$, respectively. 
This symmetric complex mass matrix is diagonalized by the
unitary $4 \times 4$ matrix $N$:
\begin{equation}
\label{eq:mixN}
N^{\ast}\mathcal{M}_N N^{\dag}\ = diag(m_{\tilde\chi_1^0},\dots,
m_{\tilde\chi_4^0}),
\hspace{2cm} 0\le m_{\tilde\chi_1^0} \le \dots \le m_{\tilde\chi_4^0} \,.
\end{equation}

\end{appendix}

\begin{figure}[H]
\begin{center}
\includegraphics{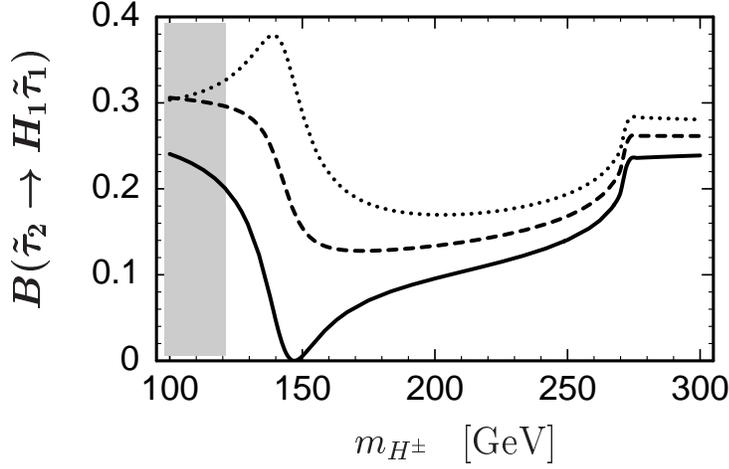}
\end{center}
\caption{Branching ratio of $\T_2 \to H_1 \T_1$, 
as a function of $m_{H^{\pm}}$ for 
$\varphi_{A_{\tau}} = 0 (\rm solid~line)$, 
$\pi/2 (\rm dashed~line)$, $\pi (\rm dotted~line)$,
$\varphi_{\mu} = \varphi_{U(1)} = 0$, 
$m_{\T_1}=240$~GeV, $m_{\T_2}=500$~GeV,
$|\mu| = 600$~GeV, $|A_{\tau}| = 900$~GeV, $\tan\beta = 30$, and
$M_2=450$~GeV, assuming $M_{\ti L} > M_{\ti E}$.
In the grey area the condition $m_{H_1} > 110$~GeV is not
fulfilled.}  
\label{fig:stau1.11}
\end{figure}

\begin{figure}[H]
\setlength{\unitlength}{1mm}
\begin{center}
\begin{picture}(150,60)
\put(-5,0){\mbox{\epsfig{figure=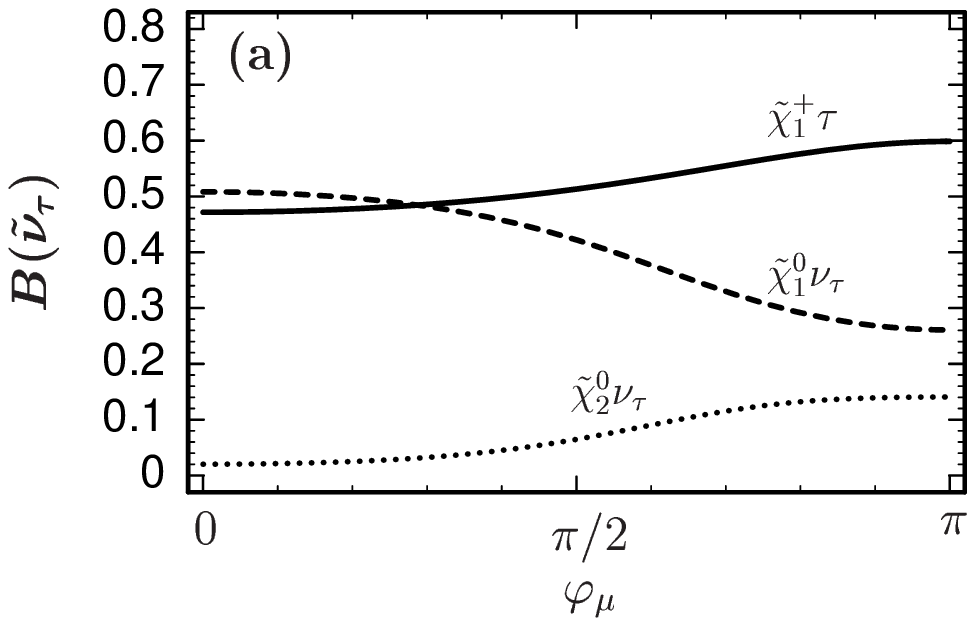,height=6.cm,width=8.cm}}}
\put(75,0){\mbox{\epsfig{figure=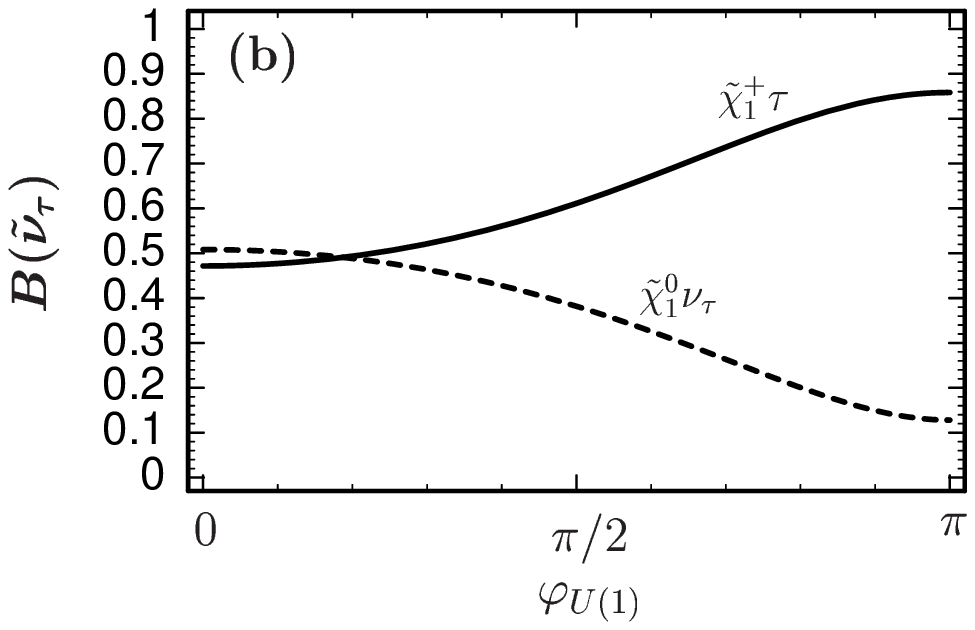,height=6.cm,width=8.cm}}}
\end{picture}
\end{center}
\caption{Branching ratios of $\SN \to \CH^0_{1,2} \nu_{\tau}$
and $\SN \to \CH^+_{1} \tau$ as a function of 
a) $\varphi_{\mu}$ for $\varphi_{U(1)}=0$ and b)
$\varphi_{U(1)}$ for $\varphi_{\mu}=0$. 
The other parameters are
$m_{\T_1}=240$~GeV, $m_{\T_2}=500$~GeV,
$M_2=500$~GeV, $|\mu|=150$~GeV, $\tan\beta=3$, 
$|A_{\tau}| = 1000$~GeV and $\varphi_{A_\tau}=0$,
assuming $M_{\ti L} < M_{\ti E}$.} 
\label{fig:stau1.8}
\end{figure}

\begin{figure}[H]
\setlength{\unitlength}{1mm}
\begin{picture}(160,180)
\put(-12,-85){\mbox{\epsfig{figure=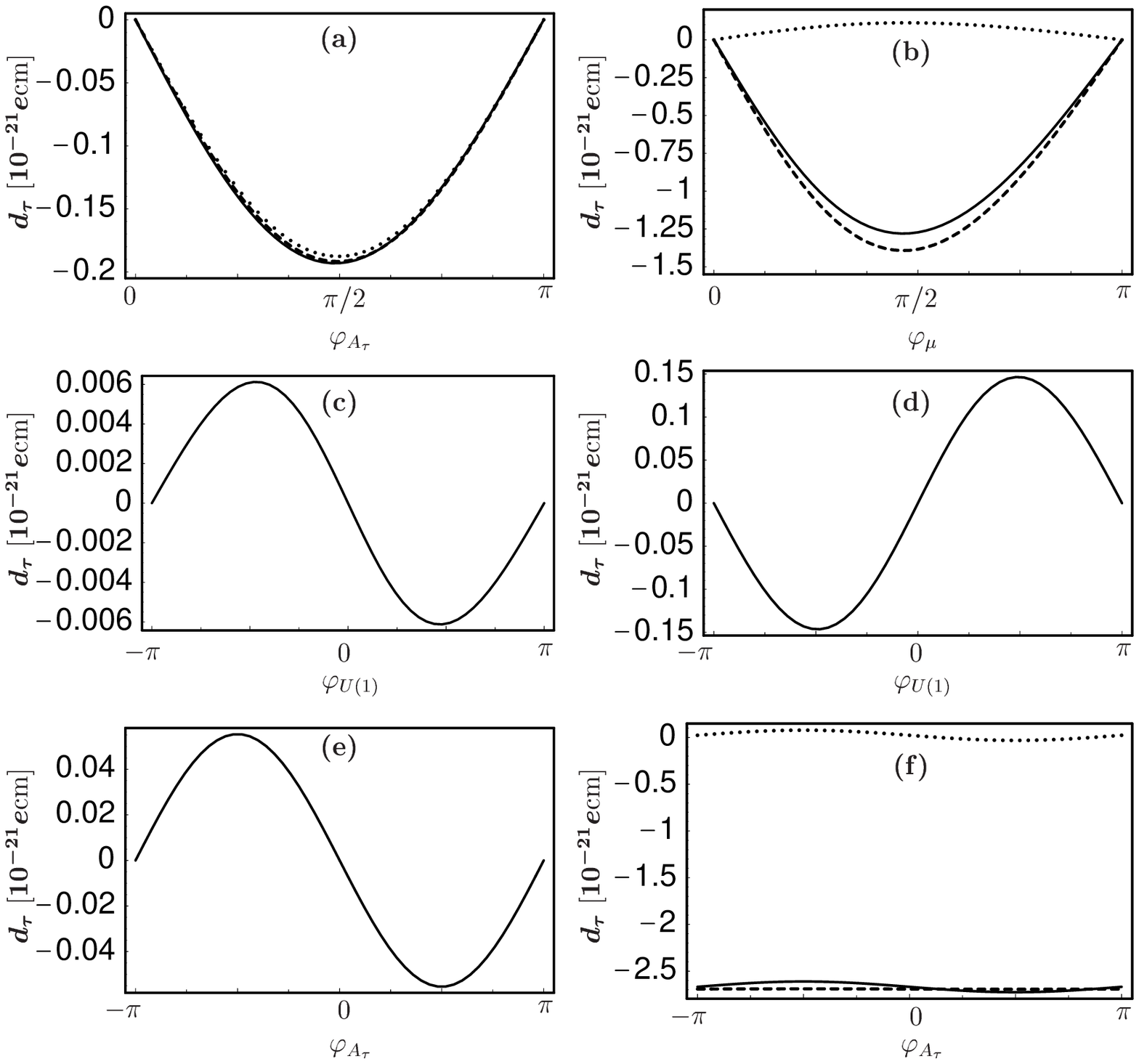,height=28.5cm,width=19cm}}}
\end{picture}
\caption{$d_\tau$ (in $10^{-21} e$cm)
  corresponding to a) Fig.~1, b) Fig.~5,
 c) Fig.~4a with $\varphi_\mu=0$, d) Fig.~4b with $\varphi_\mu=0$, 
e) Fig.~8a, and f) Fig.~8b.
 The lines in a) correspond to $m_{\SN}=233$~GeV~$(\rm solid~line)$, 
$238$~GeV~$(\rm dashed~line)$, $243$~GeV~$(\rm dotted~line)$.
 The lines in b) and f) correspond to: total $\tau$ EDM (solid line), 
 chargino--loop
 contribution (dashed line) and neutralino--loop contribution (dotted line).}
\label{fig:edm}
\end{figure}

\end {document}